\documentclass[journal, 12pt,twocolumn]{IEEEtran}
\usepackage{ifpdf}
\usepackage{cite}
\ifCLASSINFOpdf
\usepackage[pdftex]{graphicx}
\DeclareGraphicsExtensions{.pdf,.jpeg,.png}
\else
\usepackage[dvips]{graphicx}
\DeclareGraphicsExtensions{.pdf}
\fi
\usepackage{graphicx}
\ifCLASSOPTIONcompsoc
\usepackage[caption=false, font=normalsize, labelfont=sf, textfont=sf]{subfig}
\else
\usepackage[caption=false, font=footnotesize]{subfig}
\fi
\usepackage[font=small,skip=5pt]{caption}
\usepackage{amsfonts,mathtools}
\usepackage{bbm}
\usepackage{amsthm} 
\usepackage{empheq} 
\usepackage{balance}
\usepackage{epstopdf}
\usepackage{amsmath}
\usepackage{amssymb}
\usepackage{color}
\usepackage{balance}
\usepackage{array}
\usepackage{mathtools} 
\usepackage{algorithm,algorithmic} 
\usepackage{enumitem}
\usepackage[mathcal]{euscript}
\usepackage{tabularx} 
\usepackage{multirow}
\usepackage{booktabs}
\usepackage{makecell}
\usepackage{aligned-overset}
\usepackage{adjustbox}
\usepackage{lipsum} 
\usepackage{bbold}
\usepackage{url}
\usepackage[normalem]{ulem}

\newtheorem{theorem}{Theorem}

\newtheorem{lemma}{Lemma}
\newtheorem{remark}{Remark}

\newcommand{\area}{{\mathsf{C}}}
\newcommand{\ibf}[1]{{\mathit{\boldsymbol{#1}}}}

\newcommand{\device}{{\mathrm{D}}}
\newcommand{\one}{{\mathrm{[I]}}}
\newcommand{\two}{{\mathrm{[II]}}}

\begin{document}

\title{Performance Analysis of Multi-user NOMA Wireless-Powered mMTC Networks: \\ A Stochastic Geometry Approach}

\author{{Thanh-Luan~Nguyen,
		Tri~Nhu~Do,
		and Georges~Kaddoum,~\IEEEmembership{Senior~Member,~IEEE}.}%
		\thanks{T.-L. Nguyen is with the Faculty of Electronics Technology, Industrial University of Ho Chi Minh City, HCMC 700000, Vietnam (e-mail: nguyenthanhluan@iuh.edu.vn).}
		\thanks{T.~N.~Do~and~G.~Kaddoum are with the Department of Electrical Engineering, the \'{E}cole de Technologie Sup\'{e}rieure (\'{E}TS), Universit\'{e} du Qu\'{e}bec, Montr\'{e}al, QC H3C 1K3, Canada (emails: tri-nhu.do@etsmtl.ca, georges.kaddoum@etsmtl.ca).}}
		
\maketitle

\begin{abstract} 
In this paper, we aim to improve the connectivity, scalability, and energy efficiency of machine-type communication (MTC) networks with different types of MTC devices (MTCDs), namely Type-I and Type-II MTCDs, which have different communication purposes.
    To this end, we propose two transmission schemes called connectivity-oriented machine-type communication (CoM) and quality-oriented machine-type communication (QoM), which take into account the stochastic geometry-based deployment and the random active/inactive status of MTCDs. 
Specifically, in the proposed schemes, the active Type-I MTCDs operate using a novel Bernoulli random process-based simultaneous wireless information and power transfer (SWIPT) architecture.
    Next, utilizing multi-user power-domain non-orthogonal multiple access (PD-NOMA), each active Type-I MTCD can simultaneously communicate with another Type-I MTCD and a scalable number of Type-II MTCDs.
In the performance analysis of the proposed schemes, we prove that the true distribution of the received power at a Type-II MTCD in the QoM scheme can be approximated by the Singh-Maddala distribution. 
    Exploiting this unique statistical finding, we derive approximate closed-form expressions for the outage probability (OP) and sum-throughput of massive MTC (mMTC) networks.
Through numerical results, we show that the proposed schemes provide a considerable sum-throughput gain over conventional mMTC networks.
\end{abstract}


\section{Introduction}
\label{sec:introduction}

Machine-type communication (MTC) has been a key enabler of the Third Generation Partnership Project (3GPP)'s Release 16 New Radio (NR) \cite{BaekMCOM2021}, and will continue to be an essential component of the sixth generation (6G) of wireless networks \cite{Mahmood6GSUMMIT2020}.  
    For instance, the deployment of MTC for low-power applications enables long-distance communications with relaxed requirements on data rate and latency between the core network and a large number of battery-powered devices \cite{DingJIOT2021, YangMWC2017}.

    MTC networks that consist of a large number of MTC devices (MTCDs) are referred to as massive MTC (mMTC) \cite{BockelmannMCOM2016}.
MTCDs can be categorized into two classes based on their demands and characteristics \cite{LiesegangTCOMM2021}, namely one that operates continuously and has deterministic traffic demands, and another that operates in a probabilistic manner has an event-driven appearance and has sporadic traffic demands.
    To handle the coexistence of different classes of MTCDs in the same mMTC network, group-based mMTC networks, in which diverse MTCDs are divided into groups based on different criteria, such as location, delay, and traffic characteristics, have been proposed \cite{WangTCOMM2021, ChoiLWC2020}.

In this paper, we focus on five major mMTC issues. First, as pointed out in \cite{MohammadkarimiMVT2018, ZhengWC2014, ChenCST2017, SinghAccess2021, ElbayoumiCST2020}, the practical deployment and implementation of mMTC networks on a large scale faces a number of challenges due to the intensive communications between MTCDs, including congestion, coverage, energy self-sufficiency, scalability, and modelling. 
    Consequently, advanced multiple access techniques, such as non-orthogonal multiple access (NOMA), have been used to alleviate congestion caused by massive signalling overhead when a large number of MTCDs are attempting to access the spectrum  \cite{MohammadkarimiMVT2018, SinghAccess2021, ElbayoumiCST2020}.
Second, because MTCDs rely heavily on short transmissions for instantaneous communication, the network sum-throughput can be drastically reduced without efficient resource allocation algorithms.
    To overcome this barrier, group-based machine-to-machine communications, i.e., MTC according to 3GPP standards \cite{Cisco2016}, can be used for effective radio resource management and spectrum sharing, resulting in increased network throughput.
Third, in order to establish communication, MTCDs can use up their limited energy, which can be harmful to mMTC networks \cite{ElbayoumiCST2020}. 
    Therefore, wireless-powered MTC networks have been proposed in the literature as a solution to this problem \cite{QiTSP2020, TangJSTSP2019, MohammadkarimiMVT2018, LiJIOT2021}. 
    However, as the number of devices grows exponentially, a more powerful wireless-powered system designed specifically for MTC is required.
Fourth, because of the ubiquitous coverage of MTCDs, their locations are typically considered random \cite{ChenCST2017}, making modelling the spatial distribution of the MTCDs deterministic, unappealing, and unscalable.
    A few recent works on stochastic mMTC networks propose modelling the deployment of mMTC using stochastic geometry as a solution to this critical problem \cite{KamelJIOT2020, LiesegangSPAWC2020, LiuIoT2021}.
Fifth, transmissions between distant MTCDs may suffer from severe data losses due to multipath fading and intra-interference from other MTC devices, eventually reducing the network throughput.
    To address this issue, deploying several regenerative connections among the MTCDs in a multi-hop fashion can increase network coverage and, as a result, network sum-throughput and connectivity.

\subsection{Novel Communication Strategies for Diverse, Group-based mMTC Networks}
To address the aforementioned issues and improve the scalability of MTC networks and energy efficiency at the MTCDs, we propose two transmission schemes as follows. 
    In the case of a high power budget, we propose the connectivity-oriented machine-type communication (CoM) scheme, which serves all Type-II MTCDs in the vicinity, and in the case of a low power budget, the quality-oriented machine-type communication (QoM) scheme, which opportunistically serves a single Type-II MTCD.
 Specifically, we consider two types of MTCDs as follows: MTCDs with seamless operation in a multi-operator network \cite{Mahmood6GSUMMIT2020}, called Type-I MTCDs, which are able to store-and-forward messages through a series of regenerative relaying transmissions; MTCDs with consistent connectivity \cite{Mahmood6GSUMMIT2020}, called Type-II MTCDs, which operate under stringent instantaneous QoS requirements.
    Notably, we define the random active and/or inactive status of MTCDs as follows: the locations of active or inactive Type-II MTCDs form a homogeneous Poisson point process (HPPP), and their number within a close vicinity of each Type-I MTCD is Poisson distributed, as will be explained in detail in the following section.
    Following that, we propose Bernoulli-based energy harvesting (BEH) protocols, in which energy harvesting is performed based on the demand of Type-I MTCDs. 
In addition, we use multi-user power-domain NOMA (PD-NOMA) to improve connectivity by allowing a Type-I MTCD to serve other Type-I and Type-II MTCDs at the same time.
    In particular, we use stochastic geometry (SG) to capture the random spatial distribution of a set of active and inactive MTCDs to group MTCDs in this phase.

Recently, downlink mMTC and downlink mMTC-NOMA have been gaining much attention from the research community \cite{MishraIoT2021, SoniICTC2021, GoktasLWC2022, MishraCL2022}. 
    The authors in \cite{MishraIoT2021} proposed a joint subcarrier and power allocation strategy to maximize the number of connected devices in IoT network consisting of MTCDs using NOMA in the downlink. 
Recently, the authors in \cite{GoktasLWC2022} proposed a wireless power transfer-assisted NOMA (WPT-NOMA) transmission scheme for mMTC.
    The authors of \cite{MishraCL2022} integrated rate-splitting multiple-access (RSMA) with time-division-duplex Cell-free massive MIMO for downlink mMTC.
     In addition, it is envisioned that envisioned that mMTC can be implemented in millimeter Wave (mmWave)/Terahertz (THz) wideband in 6th Generation (6G) wireless networks \cite{ShaoTWC2022}.
As a result, a distributed multi-hop relaying is likely to be practical for such networks. 
    In particular, the authors in \cite{IbrahimTC2021} proposed routing techniques for mmWave networks, namely minimum hop count (MHC) and nearest Line-of-Sight (LoS) relay to the destination with MHC (NLR-MHC).
It is noted that \cite{SoniICTC2021} has introduced a multi-hop mMTC decode-and-forward (DF) relaying system with the digital enhanced cordless telecommunication new radio (DECT-2020 NR) standard.
    However, regenerative mMTC-NOMA with multi-hop infrastructure has not been given out in literature. As a result, its theoretical work on the outage probability (OP) and throughput analysis are still open issues.

\subsection{Related Works} 

In this subsection, we elaborate on recent related works on NOMA, simultaneous wireless information and power transfer (SWIPT), and their applications to MTC/mMTC networks.
    {In the literature, WPT has been considered for mMTC. For instance, the WPT-NOMA offers an alternative for energy-constrained devices \cite{AzmatCL2016, CayamcelaAccess2019, GoktasLWC2022}. 
The authors in \cite{AzmatCL2016, CayamcelaAccess2019} proposed integrating the MTCDs with WPT technologies.
In particular, energy harvesting (EH) from ambient radio frequency (RF) sources is utilized for prolonging the lifespan of energy-constrained MTCDs.
    In \cite{GoktasLWC2022}, a WPT-NOMA transmission scheme for mMTC networks is proposed for uplink and downlink communications. 
The proposed WPT-NOMA enables MTCDs to simultaneous EH and information transmission, which is crucial for the support of massive number of energy-constrained MTCDs.}
    For high connectivity, yet energy-efficient and self-sustainable mMTC networks, SWIPT-NOMA has been considered as a key enabling technology \cite{QiTSP2020}. 
The combination of SWIPT and NOMA (SWIPT-NOMA) is a potential solution to improve the spectral and energy efficiency (EE) of fifth-generation (5G) networks and beyond, especially in order to support the functionalities of the Internet-of-Things (IoT) and MTC scenarios \cite{TangJSTSP2019}. 
    Specifically, PD-NOMA has been shown to be a promising technology for improving the connectivity of MTC networks \cite{MohammadkarimiMVT2018}. 
The outage performance of relay selection strategies in SWIPT-NOMA under the assumption of residual hardware impairments and channel estimation errors was studied in \cite{LiJIOT2021}. 

In \cite{ChoiLWC2020}, the authors focused on the diverse and scalable connectivity of MTCDs by proposing two distinctive types of devices in MTC networks.
    In the case of different types of MTCDs coexisting in the same network, the authors in \cite{QiaoTWC2021} proposed sophisticated scheduling algorithms to set up connections for different active MTCDs.
Recently, random access with layered preambles (RALP) has been proposed to support devices with two different priorities based on the concept of PD-NOMA \cite{ChoiTWC2021}.
	In addition, the work in \cite{BockelmannMCOM2016} introduced PHY and MAC layer solutions developed {for {\em the mobile and wireless communications enablers for the twenty-twenty information society} (METIS) project} to address the scalability and efficient connectivity of a massive number of devices sending very short packets.
Moreover, the authors in \cite{WangTCOMM2020} studied the application of NOMA in MTCDs to increase network connectivity. Particle swarm optimization (PSO) was implemented to improve the sum-throughput of the proposed group-based MTCD-NOMA networks.

    To enhance the coverage of MTC networks, the authors in \cite{JiACCESS2019} proposed a multi-hop cooperative transmission where signaling interaction mechanisms were developed and the relevant structure of frames was modified to be compatible with current Wi-Fi architectures.
In \cite{AtatTCOMM2017}, the authors used cooperative relaying-based device-to-device (D2D) communications to improve the transmission range and coverage of MTC networks.
    In \cite{SwainTMC2017}, considering the coverage and average rate analysis of multi-hop MTC networks using stochastic geometry, the authors showed that multi-hop D2D communication is able to increase the coverage and average rate of MTC networks. Note that the usage of NOMA was overlooked in \cite{SwainTMC2017} and \cite{AtatTCOMM2017}.
In \cite{WangACCESS2019}, the authors proposed multi-hop transmission to enhance the scalability and energy efficiency of wireless software-defined-networking (SDN)-based mMTC networks. 
    Recently, in \cite{ChampatiTNET2020}, the authors investigated the transient behavior of multi-hop wireless transmission in MTC networks under static routing and showed that such a multi-hop scheme can be applied to mMTC.

For vast numbers of devices in mMTC networks, deterministic pathloss models fail to capture the spatial randomness of the MTCDs. 
    Stochastic geometry (SG) is an important mathematical tool used to capture the spatial distribution of these devices.
In \cite{AtatTCOMM2017} and \cite{KamelJIOT2020}, HPPPs were adopted to model the network density and the MTCDs' locations. 
    Specifically, the work in \cite{AtatTCOMM2017} studied the joint spectral efficiency and fairness in an underlay hybrid SG-based network consisting of MTC, D2D-enabled, and cellular devices.
Moreover, the work in \cite{KamelJIOT2020} considered an SG-based uplink NOMA ultra-dense network consisting of human-type communication (HTC) and MTC devices. 
    In this work, the authors aimed to ensure connectivity and maintain quality-of-service for a massive number of MTCDs. 
The results showed a significant increase in network throughput due to the large number of connections from MTCDs.
    Unlike the above works, the authors in \cite{LiesegangSPAWC2020} used Mat\'ern cluster processes to model the wireless-powered MTC networks. 

\subsection{Contributions of the Paper}
The main challenges in the performance analysis of the proposed transmission schemes, which are the QoM and CoM schemes with EH at Type-I MTCDs, are highlighted as follows. 
    In the considered multi-user PD-NOMA wireless-powered mMTC network, determining exact formulas for the end-to-end (e2e) outage probability (OP) can be an intricate task.
This is mainly because the e2e OP of each individual Type-I MTCD transmission is heavily dependent on the harvested power and the performance of the previous hops. 
    Specifically, as the number of traversed regenerative connections increases, the complexity of the analytical derivations also becomes noticeable.
Second, evaluating the performance of mMTC networks can be challenging. 
    This is mainly because the performance of an MTCD is determined by the device's QoS requirements. 
As a result, obtaining the instantaneous QoS requirements of all MTCDs is essential for performance analysis, which becomes increasingly challenging as the number of MTCDs grows.
    The key contributions of our paper can be summarized as follows:
\begin{itemize}
	\item To address the aforementioned problems associated to mMTC networks, we propose two novel transmission schemes, namely the CoM and QoM schemes, which have not been reported in the literature. 
	\item Moreover, we consider the random active/inactive status of Type-II MTCDs and their stochastic geometry-based locations, which reflects a practical realization of mMTC networks.
	\item To investigate the performance of the proposed schemes, we provide closed-form expressions for the network OP, sum-throughput, and maximum reliable transmission rate for the proposed schemes.
	Specifically, the total probability theorem is adopted not only to relax the recursive formulations of the transmission power from Type-I MTCDs, but also to tackle the joint statistics of multiple non-mutually-exclusive events. 
	The results are then expressed in terms of Meijer G-functions and Fox H-functions, where the Singh-Maddala distributions are adopted to assist the analysis of QoM-based schemes.
	The obtained closed-form expressions are validated through simulations, which provide technical insights into the performance of the proposed schemes in mMTC networks. 
	\item Through numerical results\footnote{It is noted that our results are reproducible using our Matlab code, which is available at \url{https://github.com/thanhluannguyen/mMTC-WPT-NOMA} }, we obtain new findings for the considered mMTC networks as follows. By using either the CoM or QoM schemes, the network sum-throughput of Type-I and Type-II MTCDs increases, whereas the outage performance of Type-I MTCDs decreases. Thus, there is a trade-off between the achievable performance of each type of MTCD; i.e., the network performance improvement comes at the cost of some degradation for Type-I MTCDs.
	\item In addition, the maximum reliable transmission rate (MRTR), which is a function of the PD-NOMA power allocation coefficients and the probabilistic EH splitting parameters, is the key factor for balancing the performance trade-off between different types of MTCDs in the network.
	\item Finally, the implementation of PD-NOMA in multi-hop-based networks can cause an outage performance degradation by allocating a portion of the transmission power to serving extra MTCDs instead of utilizing full power for relaying information.
    On the other hand, by enabling additional connections, the overall network connectivity and sum-throughput can be effectively improved.
    It is noted that SG-based NOMA networks heavily rely on the network density and the spatial separations among the paired devices. 
    As a result, even when numerous Type-II MTCDs are available, Type-I MTCDs only schedule those that are suitably distanced apart to execute NOMA transmissions.
\end{itemize}

The rest of this paper is organized as follows: Section II presents the system model including the HPPP-based deployment of the MTCDs and the fading channel model. In Section III, we introduce the proposed probabilistic EH architectures and the proposed transmission schemes. The performance analysis of the proposed schemes in terms of OP and throughput are presented in Section IV. Section V shows the asymptotic analysis. Section VI presents the numerical results. Finally, Section VII concludes this work.

\noindent\textbf{Notations:} $\mathcal{CN}(\mu, \sigma^2)$ denotes the circular complex Gaussian random variable with mean $\mu$ and variance $\sigma^2$, $\gamma(a,x)$ denotes the lower incomplete Gamma function \cite[Eq. (8.350.1)]{Gradshteyn2007}. In addition, ${\bf 0}_v$ and ${\bf 1}_v$ denote vectors containing $v$ elements of zeros and ones, respectively. Moreover, $\Pr[\cdot]$ denotes the probability operator, $F_X(x) = \Pr[X < x]$, $\bar{F}_X(x) \triangleq \Pr[ X \ge x ]$, and $f_X(x)$ are the cumulative distribution function (CDF), the complementary CDF (CCDF), and the probability density function (PDF) of the random variable (RV) $X$.

\section{System Model}

\begin{figure*}
\centering
    \includegraphics[width=0.8\linewidth]{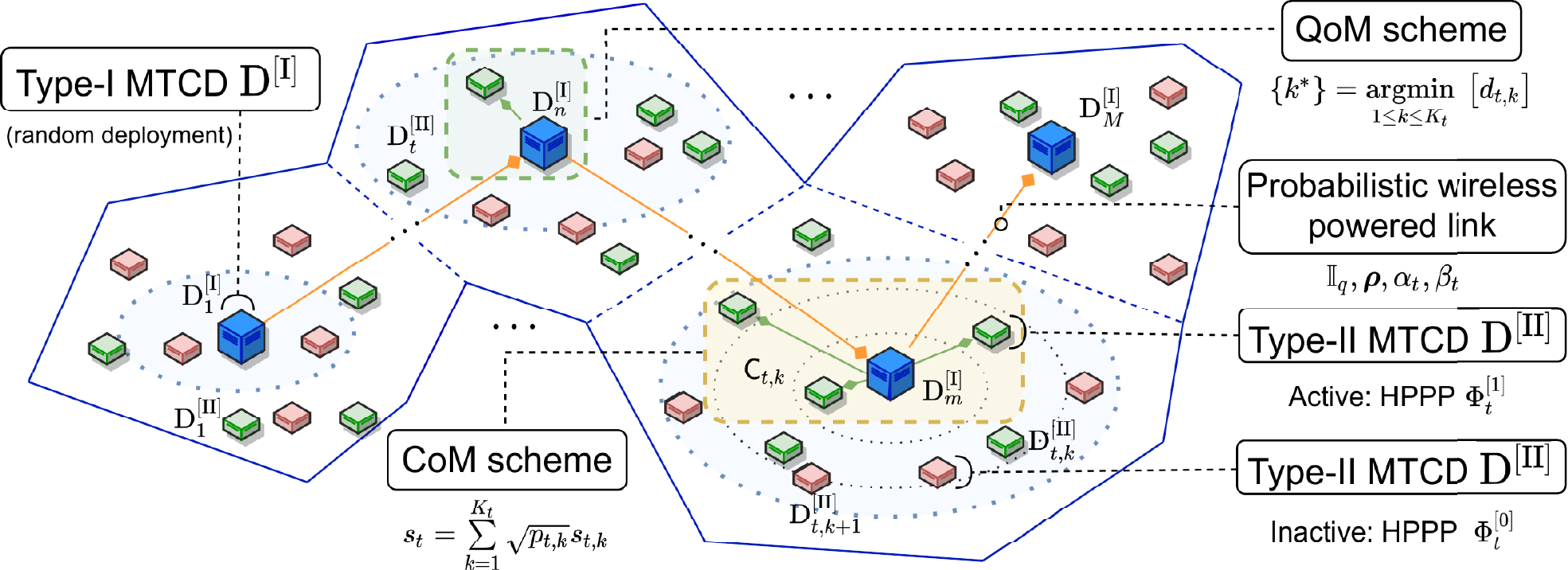} 
	\caption{Illustration of the proposed multi-user PD-NOMA probabilistic wireless-powered mMTC networks with stochastic deployment of MTC devices.}
	\label{fig1}
\end{figure*}

Let us consider an mMTC network, where a Type-I MTCD $\device^\one_1$ plans to communicate with the Type-I MTCD $\device^\one_{M}$, as depicted in Fig. \ref{fig1}. 
    Due to spatial separation, $\device^\one_1$ requires the help of other Type-I MTCDs, namely $\device^\one_2$, $\device^\one_3$, $\dots$, $\device^\one_{M-1}$, to store-and-forward the intended message to $\device^\one_M$ via multi-hop regenerative relaying transmission. 
In the proposed network, there coexists a massive number of active and inactive Type-II MTCDs, as depicted in Fig. \ref{fig1}. 
{Specifically, Type-II MTCDs can be viewed as IoT devices that are non-stationary, and stochastically spatially distributed around Type-I MTCDs; in addition, a Type-I MTCDs can be viewed as an central node that connect other Type-II MTCDs by means of access via an aggregation node or short-range device-to-device access, as considered in \cite{CayamcelaAccess2019}.}
    To enhance the connection efficiency of the considered network, a portion of the active Type-II MTCDs are served by Type-I MTCDs. 
Specifically, during the $t$-th time slot, the corresponding Type-I MTCD, denoted as $\device^{\one}_t$, can employ PD-NOMA to serve up to $K_t$ active Type-II MTCDs situated nearby, in addition to relaying information to $\device^{\one}_{t+1}$, where $1 \le t \le M-1$.
    In addition, each $\device^\one_{t+1}$ can probabilistically perform energy harvesting from the information forwarded by $\device^\one_{t}$ and exploit the harvested power for information transmission during the subsequent time slot.
The EH behavior of $\device^\one_{t+1}$ can be characterized by a Bernoulli RV, as detailed in section \ref{section_proposed_schemes}. 
    Hereafter, let us use $\device^\one$ and $\device^\two$ as acronyms for Type-I and Type-II MTCDs, respectively, and $\device^\two_{t,k}$ to refer to the $k$-th paired $\device^\two$ during the $t$-th time slot.

\subsection{Homogeneous Poisson Point Process-based mMTC Network Deployment}
Considering a square area of $x \times x$ [m$^2$] in a two-dimensional (2D) Cartesian coordinate system, $\device^\one$s are randomly distributed over the network area, and their locations are assumed to be fixed over the transmission from one $\device^\one$ to another $\device^\one$. 
    Over the circular coverage area $\area_t$ with radius $r_t$ around $\device^\one_t$, the active and inactive $\device^\two$s are modeled by HPPPs $\Phi_t^{[1]}$ and $\Phi_t^{[0]}$ {with densities per meter square} $\lambda_t^{[1]}$ and $\lambda_t^{[0]}$, respectively.
Thus, the number of active and inactive Type-II MTCDs within the proximity of $\device^\one_t$ are modeled by Poisson distributions with mean intensities $\pi \lambda_t^{[1]} r_t^2$ and $\pi \lambda_t^{[0]} r_t^2$, respectively.
    As a result, the null probability of the 2D Poisson process $\Phi^{[i]}_t$ can be expressed as
\begin{align}
\Pr\left[ |\Phi^{[i]}_t| = 0 \right] = e^{ -\lambda_t^{[i]} \pi r_t^2 },
\label{eq_iota_m}
\end{align}
where $i \in \{ 1,0 \}$.
Next, the coverage area $\area_t$ can be equally divided into $K_t$ disjoint subareas, namely $\area_{t,1}$, $\area_{t,2}$, $\dots$, $\area_{t,K_t}$. 
    Each area represents an annulus-shaped area around $\device^\one_t$ with inner and outer radiuses of $ \frac{k-1}{K_t} r_t $ and $\frac{k}{K_t} r_t$, respectively.
Accordingly, $\device^\two$s within each $\area_{t,k}$ construct a thinned HPPP ($\Phi^{[i]}_{t,k}$) from the independent thinning of the parent HPPP ($\Phi^{[i]}_t$) where the thinning probability is $\frac{1}{K_t}$. 
Hence, the average number of $\device^\two$s within $\area_{t,k}$ follows Poisson distribution with intensity $\frac{\lambda^{[i]}_t}{K_t} \pi r_t^2$.
    The null probability of $\Phi^{[i]}_{t,k}$ can then be expressed~by
\begin{align}
\Pr\left[ |\Phi^{[i]}_{t,k}| = 0 \right] = e^{ -\frac{\lambda^{[i]}_t}{K_t} \pi r_t^2 }.
\end{align}
It is noted that the number of active or inactive Type-II MTCDs is different from one annulus-shaped area to another due to the density $\lambda^{[i]}_t$.

\subsection{Fading Channel Statistics}
In order to capture the propagation effects, let us employ large-scale and small-scale models as follows. 
    Let ${\bf x}_m \in \mathbb{R}^2$ denote the locations of $\device^\one_m$, where $1 \le m \le M$, and $l_{q} \triangleq |{\bf x}_{q-1} - {\bf x}_{q}|$ be the distance from $\device^\one_{q-1}$ to $\device^\one_{q}$, where $2 \le q \le M$.
Let ${\bf x}_{t,k} \in \mathbb{R}^2$ be the location of the $k$-th selected active $\device^\two$ within the coverage area $\area_{t}$, i.e., $\device^\two_{t,k}$, thus $d_{t,k} \triangleq |{\bf x}_{t}-{\bf x}_{t,k}|$ specifies the distance from $\device^\one_{t}$ to $\device^\two_{t,k}$.
    Accordingly, the corresponding large-scale path-loss models are denoted by ${\ell}(l_m)$ and ${\ell}(d_{t,k})$. 
Specifically, for the e2e transmissions, we adopt the non-line-of-sight (NLoS) condition for the 3GPP Urban Micro (UMi) path-loss model \cite{3GPP,BjornsonWCL2020} as ${\ell}(x) = - G_r - G_t + 22.7 + 26\log_{10}(f_c) - 36.7\log_{10}(x/d_0)$ [dB], where ${\ell}(x) = \mathcal{L}( \frac{d_0}{x} )^{\epsilon}$ in linear scale with $G_r$ and $G_t$ being the receiving and transmitting antenna gains, respectively, $f_c$ (GHz) denotes the carrier frequency, $\epsilon = 3.67$, $d_0$ is the reference distance in meter and
\begin{align}
\mathcal{L} = 186.2087 \frac{(f_c)^{2.6}}{10^{0.1(G_r+G_t)}}.
\end{align}
Based on the spatial HPPP models, the Type-II MTCDs are uniformly distributed within the annulus $\area_{t,k}$ \cite{AtatTCOMM2017}, \cite{KamelJIOT2020}. Accordingly, the PDF of the distance from $\device^{\one}_t$ to each $\device^{\one}_{t,k}$ within $\area_{t,k}$, denoted by $f_{d_{t,k}}(r)$, can be expressed as
\begin{align}
f_{d_{t,k}}(r)
	= \frac{2 K_t r}{r_t^2} 
		\bigg[ \zeta \Big( r-\frac{k-1}{K_t} r_t \Big) - \zeta \Big( r-\frac{k}{K_t} r_t \Big) \bigg],
\end{align}
where $r > 0$ and $\zeta(\cdot)$ is the unit step function.
	
For the small-scale propagation, we assume that all channels experience Rayleigh fading, where the channel coefficients from $\device^\one_{q-1}$ to $\device^\one_{q}$ and from $\device^\one_{t}$ to $\device^\two_{t,k}$ are denoted as ${h_q \sim \mathcal{CN}(0,1)}$ and ${g_{t,k} \sim \mathcal{CN}(0,1)}$, respectively.
    Subsequently, the CDF of the joint large-scale and small-scale gain, denoted as ${\phi_q \triangleq |h_q|^2 {\ell}(d_q)}$, can be expressed as ${F_{\phi_q}(\phi) = 1-\exp\big[ -\phi/{\ell}(d_q) \big]}$, $\phi > 0$. 
In addition, since the active Type-II MTCDs within $\area_{t,k}$ are uniformly distributed, ${\vert g_{t,k} \vert^2 {\ell}( d_{t,k} ) \triangleq \varphi_{t,k}}$ are mutually independent and identically distributed following the CDF

{\allowdisplaybreaks
\begin{align}
&F_{\varphi_{t,k}}(\varphi) 
	= 1 - \frac{2}{\epsilon}
	\frac{ ( \frac{\varphi}{\ell(r_t)} )^{-\frac{2}{\epsilon}} }{ (\frac{k}{K_t})^2 - (\frac{k-1}{K_t})^2 }
\nonumber\\&
\times
	\bigg[
		\gamma\left(
			\frac{2}{\epsilon}, \frac{\varphi}{\ell(r_t)} 
			\left( \frac{k}{K_t} \right)^\epsilon
		\right) 
\nonumber\\&
\qquad\qquad
	    - \gamma\left(
			\frac{2}{\epsilon}, \frac{\varphi}{\ell(r_t)} 
			\left( \frac{k-1}{K_t} \right)^\epsilon
		\right)
	\bigg],\varphi > 0.\!
\end{align}}

\section{The Proposed Multi-user NOMA Probabilistic wireless-powered mMTC Schemes} \label{section_proposed_schemes}

We propose Bernoulli random process-based energy harvesting architectures to power the Type-II MTCDs. 
    {It is worth noting that the data arrive at the devices via a Bernoulli arrival process \cite{Robertazzi2000}. In heavy traffic, devices use their own energy to process heavy data, whereas in light traffic, devices employ EH to preserve their own energy. Therefore, in our proposed transmission schemes, we characterize the EH activity as a Bernoulli process.}
In addition, considering PD-NOMA, we proposed some transmission schemes for the considered mMTC networks.
    In our proposed EH architectures, each $\device^\one_{t+1}$ may or may not perform EH during the $t$-th time slot.
Let $\rho_{t+1}$ denote the probability that $\device^\one_{t+1}$ performs EH where the Bernoulli trial ${\mathbbm{I}_{t+1} \in \{1,0\}}$ indicates the on/off EH status of $\device^\one_{t+1}$. 
    The value of $\mathbbm{I}_{t+1}$ characterizes the probabilistic EH operation of $\device^\one_{t+1}$.
Specifically, for $\mathbbm{I}_{t+1} = 0$, $\device^\one_{t+1}$ will likely not perform EH with probability $1-\rho_{t+1}$; it will prefer to consume its inner battery to transmit with a fixed transmission power $P_0$. 
Meanwhile,
    $\big[ \mathbbm{I}_{t+1} = 1 \big]$ implies that $\device^\one_{t+1}$ is likely to perform EH with probability $\rho_{t+1}$.
It is noted that $\device^\one_1$ and $\device^\one_M$ do not need to perform EH. 
Such probabilistic EH behavior can be modeled via a finite Bernoulli process ${ \Xi(\ibf{\rho}) = \{ \mathbbm{I}_{1},\dots,\mathbbm{I}_{t+1},\dots,\mathbbm{I}_{M} \} }$ with parameter $ \ibf{\rho} \triangleq [\rho_1,\dots,\rho_M]$, where $\rho_{t+1} \triangleq \mathbbm{E}[\mathbbm{I}_{t+1}]$.

\subsection{Bernoulli Random Process-based Time-Switching Energy Harvesting (BTEH) Architecture}

During the $t$-th time slot, the corresponding $\device^\one_{t+1}$ harvests energy from $\device^\one_{t}$ for a duration of ${\frac{T}{M} \alpha_{t+1} \mathbbm{I}_{t+1}}$ block time, where ${0 \le \alpha_{t+1} \le 1}$ is the so-called time-switching or EH ratio, and $T$ is the block time in which an information block is transmitted from $\device^\one_1$ to $\device^\one_M$.
    Accordingly, the remaining block time, ${\frac{(1-\alpha_{t+1} \mathbbm{I}_{t+1})T }{M}}$, is then consumed for information transmission from $\device^\one_{t}$ to $\device^\one_{t+1}$.
Furthermore, by implementing NOMA, $\device^\one_{t}$ can simultaneously serve $\device^\two$s while forwarding information to $\device^\one_{t+1}$.
Accordingly, the harvested energy at $\device^\one_{t+1}$ is expressed as
\begin{align}
E_{t+1} = \eta_{t+1} (P_{t} \phi_{t+1}) \frac{ \alpha_{t+1} T}{M-1} \mathbbm{I}_{t+1}, 
\label{eq_harv_Energy}
\end{align}
where $0 \le \eta_{t+1} \le 1$ denotes the energy conversion efficiency, $P_t$ denotes the transmission power of $\device^\one_t$, and $P_0$ is the fixed transmission power. 
    Suppose that all the harvested energy is consumed by each source node, the harvested power at $\device^\one_{t+1}$ can be expressed~as
\begin{align}
P_{t+1} &= (M-1) \frac{E_{t+1}}{(1-\alpha_{t+1} \mathbbm{I}_{t+1}) T} 
\nonumber\\
	&= (M-1)\frac{\alpha_{t+1} \eta_{t+1}}{1-\alpha_{t+1}} \phi_{t+1} P_{t} \mathbbm{I}_{t+1}.
\label{eq_Pm_BTEH}
\end{align}

\subsection{Bernoulli Random Process-based Power-Splitting Energy Harvesting (BPEH) Architecture}
During the $t$-th time slot, a proportion of the received power, $\beta_{t+1} \mathbbm{I}_{t+1} P_t \phi_{t+1}$, where $0 \le \beta_{t+1} \le 1$ denotes the power-splitting or the EH ratio, is utilized for EH.
    Accordingly, the remaining power, $(1-\beta_{t+1}  \mathbbm{I}_{t+1}) (P_t \phi_{t+1})$, is then exploited for information transmission. 
Similar to the BTEH architecture, the $(M-1)$-th time slot is the effective transmission duration in which the destination $\device^\one_M$ does not perform EH, and thus all received power is exploited for transmission from $\device^\one_{M-1}$ to $\device^\one_{M}$, and potentially to the downlink $\device^\two$s. 
Accordingly, the harvested energy can be expressed as
\begin{align}
    E_{t+1} = \eta_{t+1} \beta_{t+1}  \mathbbm{I}_{t+1} P_{t} \phi_{t+1} \frac{T}{M-1}.
\end{align}
Hence, the harvested power $P_{t+1}$ can be formulated~as
\begin{align}
    P_{t+1} = (M-1) \frac{E_{t+1}}{T} = \beta_{t+1} \mathbbm{I}_{t+1} \eta_{t+1} \phi_{t+1} P_{t}. 
\label{eq_Pm_BPEH}
\end{align}
It is noted that both \eqref{eq_Pm_BTEH} and \eqref{eq_Pm_BPEH} are expressed in recursive form as a function of $P_{t}$. 
    The extended expression of the transmission power during the $t$-th time slot is given in the following {\it Remark}.

\begin{remark}
\label{lem:1}
The transmission power of $\device^\one_t$, $t \in [1,M-1]$, depends on whether it adopts EH or not. Accordingly,
\begin{itemize}
    \item[a)] if $\big[ \mathbbm{I}_{t} = 0 \big]$, $\device^\one_{t}$ likely transmits with a fixed power, thus $P_t = P_0$, with probability $\rho_t$,
    \item[b)] if $\big[ \mathbbm{I}_{t} \!=\! 1 \big]$ and there exists $\tau \in [1,t-1]$ that satisfies ${\big[\mathbbm{I}_\tau \!=\! 0 \big]}$ and $\big[ \mathbbm{I}_{j} \!=\! 1 \big]$, ${\forall j \in [\tau+1,t]}$, the harvested power is
    ${P_t = P_0 \big( \Omega_{\tau+1} \phi_{\tau+1}
    	\big) \cdots \big( \Omega_{j} \phi_{j} \big) \cdots
    	\big( \Omega_{t} \phi_t \big) }$, where $\Omega_q = (M-1) \alpha_q \eta_q/(1-\alpha_q)$ for BTEH and $\Omega_q = \beta_q \eta_q$ for BPEH.
\end{itemize}
Accordingly, $P_t$ can then be formulated as
\begin{align}
P_t = 
\left\{\!\!\!
\begin{array}{cc}
\displaystyle
	P_0, & \text{if}~\big[ \mathbbm{I}_t = 0 \big], \vspace{3pt}\\
	P_0 \!\prod\limits_{i = \tau + 1}^{t}\!
		\Omega_i \phi_i, & \text{if} \left\{\!\!
			\begin{array}{cc}
				\big[ \mathbbm{I}_\tau = 0  \big], \vspace{3pt}\\
				\big[ \mathbbm{I}_j = 1 \big],~\forall j \in [\tau+1, t],
			\end{array}
			\right.
\end{array}
\right.
\label{eq_unified_P_t}
\end{align}
\end{remark}
Consider a scenario where $\rho_q = 1$, $\forall q \in [2,M]$, and therefore all $\device^\one$s, except $\device^\one_1$, perform EH with a probability of one, thus the transmission power at $\device^\one_t$ can be expressed as
\begin{align}
    P_t = P_0 \prod_{i=1}^{t}{ |h_i|^2 } 
        \prod_{i=1}^{t} { \ell(d_i) }
        \prod_{i=1}^{t}{ \Omega_i }.
\label{eq:ex:P_m}
\end{align}

\begin{remark}
\textit{
It can be observed from \eqref{eq:ex:P_m} that the harvested power $P_t$ is contaminated by the path-loss in every $i$-th time slot, where $i \in [1,t]$. This phenomenon is familiar in traditional wireless-powered regenerative relaying systems, where the path-loss from previous paths accumulates at the energy harvesting devices until, for some devices, the harvested power reaches zero, leading to a severe outage.}
\end{remark}

To tackle the above issue, one approach is to deploy power beacons within the proximity of each $\device^\one$. 
    However, this requires the power beacon within the last hops to radiate with enormous power in order to overcome the accumulation of path-loss, not to mention the high installation cost. 
Another approach is to have some $\device^\one$s, e.g., $\device^\one_\tau$ where $\tau \in [1,t-1]$, transmit with a fixed power so that the accumulated path-loss in the $i$-th hop, $\forall i \in [1,\tau]$, is eliminated in the harvested power at $\device^\one_{t+1}$, as shown in \eqref{eq_unified_P_t}.
Ideally, each $\device^\one$ should switch to energy harvesting mode if the harvested power exceeds an acceptable power threshold. 
    This also implies that there should be a probability of EH imposed on each~$\device^\one$. 

\subsection{Co-Existing Type-I and Type-II MTCDs Transmission Schemes}During the $t$-th time slot, if there exists at least one $\device^\two$ within $\area_t$ around the transmitting $\device^\one_{t}$, additional $\device^\two$s can be scheduled to jointly perform PD-NOMA.
    In this case, $\device^\one_{t}$ performs superposition coding (SC) to form a mixture of information signals dedicated to $\device^\one_M$, denoted as $s_M$, and other $\device^\two$s, denoted as $s_t$. 
Accordingly, the information signal transmitted by $\device^\one_{t}$ can be expressed as
\begin{align}
x_{t} = \sqrt{p_M} s_M + \sqrt{1-p_M} s_{t},
\end{align}
where $0 \le p_M \le 1$ denotes the power allocation for $s_M$ and $s_{t}$ specifies the mixture of the $\device^\two$s' signals.

In practice, the distance from $\device^\one_{t}$ to $\device^\one_{t+1}$ is much larger than the distance between the paired $\device^\two$s, as observed in Fig. \ref{fig1}. 
    Accordingly, to ensure transmission fairness, $\device^\one_t$ should allocate more power to $s_M$ than to $s_t$. 
In addition, if there aren't enough $\device^\two$ within the proximity of $\device^\one_t$, it then allocates all power to $s_M$.
    As a result, the achievable instantaneous rate at $\device^\one_{t+1}$ with the BTEH architecture is obtained as
\begin{align}
\mathsf{R}^{\one}_{t+1} 
	&=  \tau_{t+1}
	\log_2\left( 
		1 + \frac{ p_M P_{t} \phi_{t+1}}{(1-p_M) P_{t} \phi_{t+1} + \sigma^2}
	\right) \mathbb{M}_{t}
\nonumber\\
    &\quad+   \tau_{t+1}
    \log_2\left( 
		1 + \frac{ P_{t} \phi_{t+1}}{\sigma^2}
	\right) \mathbb{M}^{c}_{t},
\label{eq_rate_BTEH}
\end{align}
and with the BPEH architecture is obtained as
\begin{align}
&\mathsf{R}^{\one}_{t+1}
    =  \tau_{t+1}
	\log_2\left(
		1 \!+\! \frac{(1 \!-\! \beta_{t+1} \mathbbm{I}_{t+1}) P_{t} \phi_{t+1}}{\sigma^2}
	\right) \mathbb{M}^{c}_{t}
	\nonumber\\
&
    + \tau_{t+1}
	\log_2\bigg( 1 + 
\nonumber\\&\qquad
	\frac{ p_M(1 \!-\! \beta_{t+1} \mathbbm{I}_{t+1} ) P_{t} \phi_{t+1}}{ (1 \!-\! p_M) (1 \!-\! \beta_{t+1} \mathbbm{I}_{t+1}) P_{t} \phi_{t+1} + \sigma^2}
\bigg) \mathbb{M}_{t},
\label{eq_rate_BPEH}
\end{align}
where $\tau_{t+1} \triangleq \frac{1-\alpha_{t+1} \mathbbm{I}_{t+1}}{M-1}$ for BTEH and $\tau_{t+1} \triangleq \frac{1}{M-1}$ for BPEH, 
    ${\mathbb{M}_{t} \triangleq [ |\Phi^{[1]}_t| \ge 1 ]}$ is the event where at least one active $\device^\two$ exists within the proximity of $\device^\one_t$ whereas $ \mathbb{M}^{c}_{t}$ is the complementary of $\mathbb{M}_{t}$, and 
$\sigma^2$ denotes the noise power.
In order to improve the network sum-throughput, in what follows, we propose two possible device pairing techniques employed at each $\device^\one_{t}$.

\subsubsection{Connectivity-oriented Machine-Type Communication (CoM) Scheme}

In this scenario, a transmitting $\device^\one_t$ can randomly schedule $K_{t}$ active $\device^\two$s, where each $k$-th paired device is randomly distributed within $\area_{t,k}$. 
Accordingly, the transmitted power-domain multiplexed signal, $s_{t}$, can be expressed as
\begin{align}
s_{t} = \sum_{k=1}^{K_t}{ \sqrt{p_{t,k}} s_{t,k} },
\end{align}
where $s_{t,k}$ is the unit energy information signal dedicated to the randomly selected user and $p_{t,k}$ denotes the proportion of power allocated to $s_{t,k}$ which satisfies $p_M + \sum_{k=1}^{K_t}{ p_{t-1,k} } = 1$ 
and $p_{t,1} \le p_{t,2} \le \cdots \le p_{t,K_t}$ since $d_{t,1} < d_{t,2} < \cdots < d_{t,K_t}$. 

Subsequently, the scheduled $\device^\two$ first decodes the destination signal ($s_M$), and then applies successive interference cancellation (SIC) before decoding its own signals.
    Without loss of generality, $\device^\two_{t,k}$ first attempts to decode $s_{t,K_t}$ while treating the signals $s_{t,v}$, where $v \in [1,K_t-1]$, as interference.
After $s_{t,K_t}$ is successfully decoded, $\device^\two_{t,k}$ directly cancels it from the remaining received signals using SIC, $\device^\two_{t,k}$ then decodes $s_{t,K_{t}-1}$ in a similar manner. 
    This procedure is successively iterated until $s_{t,k}$ is decoded correctly \cite{MohammadkarimiMVT2018}.
Accordingly, the achievable instantaneous rates at $\device^\two_{t,k}$ for decoding $s_M$ and $s_{t,\kappa}$, where $\kappa \in [k,K_t]$, are obtained as
\begin{align}
\mathsf{R}^{\two}_{t,k} 
      &= \tau_{t+1}
      \log_2
      \left( 
      	1 + \frac{ \frac{p_M P_t}{\sigma^2} \varphi_{t,k} }{ \frac{(1-p_M) P_t}{\sigma^2} \varphi_{t,k} + 1 }
      \right),  \label{eq_gamma_m1_nk_end}
\end{align}
\begin{align}
\mathsf{R}^{\two,[\kappa]}_{t,k} 
      &= 
      \tau_{t+1}\log_2
      \left(
      	\frac{ p_{t,\kappa} \frac{P_t}{\sigma^2} \varphi_{t,k} }{ \frac{P_{t}}{\sigma^2} \varphi_{t,k} \sum_{v=1}^{\kappa-1} p_{t,v} + 1 }
      \right), \label{eq_gamma_nq_m1_nk_end}
\end{align}
respectively.

\subsubsection{Quality-oriented Machine-Type Communication (QoM) Scheme} 

In this case, to serve active critical $\device^\two$s, only the nearest active $\device^\two$ is selected among $\Phi^{[1]}_t$.
Without loss of generality, assuming that $k^\ast$ specifies the index of the scheduled $\device^\two$ such that
\begin{align}
	\{ k^\ast \} ~= \mathop{\rm argmin}\limits_{ 1 \le k \le K_t } ~ \big[ {d}_{t,k} \big],
\end{align}
the exact $t$-th hop's instantaneous rate before and after SIC at this $\device^\two$ can be expressed as
\begin{minipage}{0.47\textwidth}
	\begin{align}
    \mathsf{R}^{\two}_{M \to t} 
      &= \tau_{t+1}
      \log_2\left(
      	1 + \frac{ \frac{p_M P_{t}}{\sigma^2} \varphi_{t}}{ \frac{(1-p_M) P_{t}}{\sigma^2} \varphi_{t} + 1 }
      \right), \label{eq_RM2t}
    \end{align}
\end{minipage} \hfill
\begin{minipage}{0.49\textwidth}
	\begin{align}
    \mathsf{R}^{\two}_{t} 
     &= \tau_{t+1}
      \log_2\left(
      	1 + \frac{ (1-p_M) P_{t} \varphi_{t} }{ \sigma^2 }
      \right), \label{eq_Rt} 
    \end{align}
\end{minipage}
\\
respectively, where $\varphi_{t} \triangleq \big\vert {g}_{t,k^\ast} \big\vert^2 \ell(d_{t,k^\ast})$.

\section{Performance Analysis}

In this section, we study the outage performance of the proposed transmission schemes, which is a key performance metric for delay-limited mMTC networks.
    An outage event occurs whenever the instantaneous achievable rate drops below a pre-defined transmission rate threshold \cite{DoTCOM2021}. 
    As a result, the OP at $\device^\one$ or $\device^\two$ is defined as the probability that the outage event occurs at that device. 
    Accordingly, the e2e OP at $\device^\one$ or $\device^\two$ is defined as the probability of an outage event occurring at that device or at other devices involved in the transmission from $\device^\one_1$ to that $\device^\one$ or $\device^\two$.
In order to assist the analysis in the upcoming subsections, we introduce the following Lemma.

\begin{lemma}
\label{lemma_TypeI_BTEH_xi_tau_t}
Let us denote $\xi_{\tau,t} \triangleq \prod_{i=\tau}^{t}{ \Omega_i \phi_i }$ as the product of $(t-\tau)$ independent and not-necessarily identically distributed (i.n.i.d) exponential RVs, each with mean $\Omega_i \ell(l_i)$, the CDF and PDF of $\xi_{\tau,t}$ is obtained as

\begin{minipage}{0.47\textwidth}
	\begin{align}
        F_{\xi_{\tau,t}}(\xi)
    	&= 1 - G^{t-\tau+1,0}_{0,t-\tau+1}
    	\left[ 
        \left.
    		\frac{\xi}{ \bar{\xi}_{\tau,t} }
    	\right\vert
        	{\bf 1}_{t-\tau},0
    	\right], \label{eq_F_xi}
    \end{align}
\end{minipage} \hfill
\begin{minipage}{0.49\textwidth}
	\begin{align}
        f_{\xi_{\tau,t}}(\xi)
    	&= \frac{1}{ \bar{\xi}_{\tau,t} }
    	G^{t-\tau+1,0}_{0,t-\tau+1}
    	\left[ 
        \left.
    		\frac{\xi}{ \bar{\xi}_{\tau,t} }
        \right\vert
       		{\bf 0}_{t-\tau+1}
    	\right], \label{eq_f_xi}
    \end{align}
\end{minipage}
\\
where $\bar{\xi}_{\tau,t} \triangleq \prod_{i=\tau}^{t}{\Omega_i \ell(l_i)}$ and $G^{m,n}_{p,q}$ denotes the Meijer G-function \cite{Gradshteyn2007}.
\end{lemma}

\begin{IEEEproof}
The proof is provided in Appendix \ref{apd_a}.
\end{IEEEproof}

For brevity, let us denote ${X}_{t+1} \triangleq \frac{P_{t}}{\sigma^2} \phi_{t+1}$, ${Y}_{t,k} \triangleq \frac{P_t}{\sigma^2} \varphi_{t,k}$, and ${Z}_{t} \triangleq \frac{P_t}{\sigma^2} \varphi_t$.
    It is noted that $X_{t+1}$, $Y_{t,k}$, and $Z_{t}$ denote the received powers at the $t$-th time slot normalized by the noise power observed at $\device^\one_{t+1}$, the $k$-th $\device^\two$ in the CoM scheme,
    and the paired $\device^\two$ in the QoM scheme, respectively.
In what follows, the OP at $\device^\one_{t+1}$ during the $t$-th time slot with BTEH and BPEH architectures are given in subsection \ref{subsection_OP_type_I}.
    Accordingly, the OPs at the scheduled $\device^\two$s for ToM and CoM schemes are provided in subsection \ref{subsec_PA_typeII_MTCD}.
Using the results in subsections \ref{subsection_OP_type_I} and \ref{subsec_PA_typeII_MTCD}, we formulate the e2e OPs at the $\device^\one_M$, and each $\device^\two_{t,k^\ast}$ in the QoM scheme and each $\device^\two_{t,k}$ in the CoM scheme, where $1 \le t \le M-1$ and $1 \le k \le K_t$ in subsection \ref{subsec_SumT_mMTC}. 
    The sum-throughput of the proposed network is also provided in subsection \ref{subsec_SumT_mMTC}.
    
\subsection{Performance Analysis of Type-I MTCDs} \label{subsection_OP_type_I}

The OP of each $\device^\one_{t+1}$ is determined as the probability that the instantaneous rate drops below the target transmission rate of the destination $\device^\one_M$. 
    Furthermore, the instantaneous rate is expressed in terms of the effective transmission time.
For BTEH, the effective transmission time depends on whether $\device^\one_{t+1}$ adopts EH.
{It is noted that for a fair performance comparison, we consider the amount of power allocated for each $\device^\one$ is equivalent in both CoM and QoM schemes, the OP of $\device^\one_{t+1}$ during the $t$-th time slot is expressed as}
\begin{equation}
{\rm OP}^{\one}_{t+1} \triangleq
\Pr[ \mathsf{R}^{\one}_{t+1} < R_M ], \label{eq_OP_BXEH_DIm_SIM}
\end{equation}
where $\mathsf{R}_{t+1}^{\one}$ is given by \eqref{eq_rate_BTEH} and \eqref{eq_rate_BPEH} for BTEH and BPEH architectures, respectively, and $R_M$ denotes the target transmission rate of $s_M$, measured in bit/s/Hz. 

The outage analysis of $\device^\one$s using the BTEH architecture is obtained in the following theorems. 
    To provide a consistent comparison of the proposed transmission schemes, i.e., the CoM and QoM schemes, in the following sections, the total power allocated to $\device^\one$s in both schemes is considered to be identical.
As a result, the received information and interference powers at $\device^\one_{t+1}$ are maintained regardless of the transmission schemes adopted, resulting in similar OPs.

\begin{theorem} \label{theorem_TypeI_BTEH_anyScheme}
The closed-form expression for the OP at $\device^\one_{t+1}$, using the BTEH-powered QoM or CoM scheme, denoted as the TQoM or TCoM scheme, respectively, can be written as
\begin{align} \label{eq_OP_BTEH_Sm}
{\rm OP}^{\rm TQoM}_{\device^\one_{t+1}} &= 
{\rm OP}^{\rm TCoM}_{\device^\one_{t+1}} 
\nonumber\\
	&= 1 - \sum_{i} \sum_{j}
    {\rho}^{[i]}_{t+1} \iota^{[j]}_{t} \bar{F}_{X_{t+1}}( \tau^{[ij]}_{M} ),
\end{align} 
where for {$ \scriptstyle { \tau^{[00]}_{M} \triangleq 2^{(M-1) R_M} - 1 }$ and $ \scriptstyle { \tau^{[10]}_{M} \triangleq 2^{\frac{M-1}{1-\alpha_{t+1}} R_M} - 1 }$ }, we have
    ${ \tau^{[i1]}_{M} \!\triangleq\! \frac{ \tau^{[i0]}_{M} }{1\!-\!( \tau^{[i0]}_{M}+1)(1\!-\!p_M)} }$.
\end{theorem}

\begin{IEEEproof}
Considering the event where $\device^\one_t$ adopts EH in conjunction with a successful decoding event at $\device^\one_t$, i.e., $\mathsf{R}^{\one}_{t+1} \ge R_M$, the OP of $\device^\one_t$ during the $t$-th time slot is expressed as
\begin{align} 
&{\rm OP}^{\rm TQoM}_{\device^\one_{t+1}} = 
{\rm OP}^{\rm TCoM}_{\device^\one_{t+1}} = 1 - \sum_i \Pr[ \mathbbm{I}_{t+1} = i ] 
\nonumber \\
& \qquad\qquad\qquad\times
	\Pr[ \mathsf{R}^{\one}_{t+1} \ge R_M \mid \mathbbm{I}_{t+1} = i ].
\label{eq_OP_BTEH_proof_1}
\end{align}

It is worth mentioning that the second probability considers that BTEH is implemented at $\device^\one_{t+1}$'s receiver, and also represents the outage performance gain from performing EH during the $t$-th time slot. 
    Meanwhile, the second probability considers no EH is executed at $\device^\one_{t+1}$.
Taking into account the number of $\device^\two$s within the proximity of $\device^\one_{t}$, the first probability, denoted as ${\cal P}$, can be further expressed~as 
\begin{align} \label{eq_P1a}
&{\cal P}
	=  \Pr\big[ \mathbbm{I}_{t+1} = 0 \big]
		\Pr\big[
			X_{t+1} \ge \tau^{[01]}_{M},
			\mathbb{M}^{c}_{t}
		\big] 
\nonumber\\&+
	\Pr\big[ \mathbbm{I}_{t+1} = 0 \big]
	\!\Pr\!\bigg[
		\frac{p_M X_{t+1}}{ (1-p_M) X_{t+1} + 1} \ge \tau^{[01]}_{M},
		\mathbb{M}_{t}
	\bigg].
\end{align}

It is noted that $X_{t+1}$ is independent of $\mathbb{M}^{c}_t$ and $\mathbb{M}_t$, thus one can separate their joint probability into a product of the corresponding probabilities such that
\begin{align}
{\cal P}
	=	\rho^{[0]}_{t+1}
    \{
		{\iota}^{[1]}_{t} \Pr[ X_{t+1} \ge \tau^{[01]}_{M} ]
	    \!+\! 	{\iota}^{[0]}_{t} \Pr[ X_{t+1} \ge \tau^{[00]}_{M} ]
	\}.
\label{eq_calP_final}
\end{align}

As shown in \eqref{eq_calP_final}, both probabilities are expressed via $\bar{F}_{X_{t+1}}(x)$, i.e., the CCDF of $X_{t+1}$. 

Next, the CDF of the received power at $\device^\one_{t+1}$ normalized by the noise power in the $t$-th time slot can be expressed as
\begin{align}
\bar{F}_{X_{t+1}}(x) 
	&= \Pr[ X_{t+1} \ge x , \mathbbm{I}_t = 0 ] 
\nonumber\\
	&\qquad\qquad
	+ \Pr[ X_{t+1} \ge x , \mathbbm{I}_t = 1 ].
\label{eq_lemma_FX_sim}
\end{align}
Denoting $X_{t+1}^{[0]} \triangleq \Pr[ X_{t+1} \ge x , \mathbbm{I}_t = 0 ]$ and $X_{t+1}^{[1]} \triangleq \Pr[ X_{t+1} \ge x , \mathbbm{I}_t = 1 ]$, their CCDFs are obtained as
\begin{align}
&\bar{F}_{X_{t+1}^{[0]}}(x) =
    {\rho}^{[0]}_{t}
	\exp\left(
		- \frac{x}{\bar{\gamma}_0 \ell(l_{t+1})}
	\right), \label{eq_F_Xm_0} \\
&\bar{F}_{X_{t+1}^{[1]}}(x)
	= \sum_{\tau=1}^{t-1}
	\Bigg[
		{\rho}^{[0]}_\tau
		\prod_{\mathclap{j=\tau+1}}^{t} { \rho^{[1]}_j }
	\Bigg]
\nonumber\\&\qquad\times
    G^{t-\tau+1,0}_{0,t-\tau+1}
	\Bigg[
	\left.
		\frac{x}{ \bar{\gamma}_0 \ell(l_{t+1}) \bar{\xi}_{\tau+1,t}}
	\right\vert
		{\bf 1}_{t-\tau},0 
	\Bigg], \label{eq_F_Xm_1}
\end{align}
where $x > 0$ and $\bar{\gamma}_0 \triangleq \frac{P_0}{\sigma^2}$. The derivation of \eqref{eq_F_Xm_0} and \eqref{eq_F_Xm_1} is provided in Appendix \ref{apendix_proof_lemma_X_fin}.

By invoking \eqref{eq_F_Xm_0} and \eqref{eq_F_Xm_1}, we obtain the exact analytical expression of $\mathcal{P}$ in \eqref{eq_calP_final}. The second probability in \eqref{eq_OP_BTEH_proof_1} can then be derived in an analogous manner, i.e., replacing $\tau^{[00]}_M$ and $\tau^{[01]}_M$ with $\tau^{[10]}_M$ and $\tau^{[11]}_M$, respectively. This completes the proof of Theorem \ref{theorem_TypeI_BTEH_anyScheme}.
\end{IEEEproof}
It should be noted that $\bar{F}_{X_{t+1}}(x)$ is defined for $x > 0$, since received powers are always positive.
    Hence, $\bar{F}_{X_{t+1}}( \tau^{[01]}_{M} ) > 0$ and $\bar{F}_{X_{t+1}}( \tau^{[11]}_{M} ) > 0$ when both $\tau^{[01]}_{M}$ and $\tau^{[11]}_{M}$ are positive, which is equivalent~to
\begin{align}
    \tau^{[i1]}_{M} < \frac{p_M}{1-p_M}, \forall i \in \{0,1\}. \label{eq_a1_a2_condition} 
\end{align}
This also implies that the target transmission rate of $s_M$ so that $\bar{F}_{X_{t+1}}( \tau^{[01]}_{M} )$ and $\bar{F}_{X_{t+1}}( \tau^{[11]}_{M} )$ are theoretically positive should satisfy
\begin{align}
    R_M < \frac{1-\alpha_{t+1}}{M-1} \log_2\left( 
        1 + \frac{p_M}{1-p_M}
    \right),
\label{eq_target_rate_condition_BTEH}
\end{align}
which is obtained by solving \eqref{eq_a1_a2_condition} for $R_M$. 
    Henceforth, let the right-hand side of \eqref{eq_target_rate_condition_BTEH} be depicted as the MRTR of the $\device^\one_M$'s desired information signal using the TCoM or TQoM scheme. 
The MRTR of $s_M$ in \eqref{eq_target_rate_condition_BTEH} is undefined when $p_M = 1$, which implies 
    a) \textit{$\device^\one$s only focus on relaying information to $\device^\one_M$ and neglect the demand of $\device^\two$s, and}
    b) \textit{the interference power from the mixture of $\device^\two$s while decoding $s_{M}$ is neglected}.
    
In general, the MRTR of an information signal is defined for NOMA-aided transmission in delay-limited systems and is the minimum target transmission rate that cannot be satisfied.
    As a result, the MRTR can offer effective upper-bounds for the target transmission rate, ensuring that the performance gains from NOMA operations are theoretically greater than zero, thus improving transmission reliability.
    
\textcolor{black}{
From \eqref{eq_OP_BTEH_Sm}, one can observe that the OP of $\device^\one$ during the $t$-th time slot relies on the following key factors:
\begin{itemize}
    \item[a)] 
        Density of $\device^\two$s around each dedicated $\device^\one_t$ where $1 \le t \le M-1$: As $\lambda^{\one}_t$ increases, $\iota^{[1]}_t \to 0$. In this case, the power allocation has a significant impact on the outage performance. In contrast, at ${\iota}^{[1]}_t = 0$, the network gains no benefit from performing NOMA and the power allocation has no impact on the outage at $\device^\one_t$.
    \item[b)] 
        The EH ratio $(\alpha_{t+1})$: As $\alpha_{t+1}$ increases, $\tau^{[10]}_M$ also increases until ${\rho}^{[1]}_{t+1} \iota^{[0]}_{t} \bar{F}_{X_{t+1}}( \tau^{[10]}_{M} )$ in \eqref{eq_OP_BTEH_Sm} reaches zero and there is no benefit from performing EH. 
    	Specially, by solving $\tau^{[10]}_M = \frac{p_M}{1-p_M}$ for $\alpha_{t+1}$, we find that at $\alpha_{t+1} = \frac{\log_2(1-p_M) + (M-1) R_M}{\log_2(1-p_M)}$, the ${\rho}^{[1]}_{t+1} \iota^{[1]}_{t} \bar{F}_{X_{t+1}}( \tau^{[1]}_{M} )$ component suddenly drops to zero, which in turn leads to a drastic drop in the performance.
    \item[c)] 
        Power allocation at $\device^\one_{t}$: as pointed out in a) this factor only affects the OP when $\iota^{[1]}_t < 1$ since at $\iota^{[1]}_t = 1$, no $\device^\two$ is active within the proximity of $\device^\one_{t}$. Hence, $\device^\one_t$ should allocate all its power for relaying information to $\device^\one_{t+1}$. 
\end{itemize}}

Note that ${\rm OP}^{\rm TQoM}_{\device^\one_{t+1}}$ and ${\rm OP}^{\rm TCoM}_{\device^\one_{t+1}}$ also depend on other factors in of $\iota^{[1]}_{t}$ and the derived CDF of $X_{t+1}$, such as the coverage area, the distance between two consecutive $\device^{\one}$s, $\eta_t$, $\epsilon$, etc. However, the above key factors are obvious just by observing the expression derived in \eqref{eq_OP_BTEH_Sm}.
\begin{theorem}
\label{theorem_TypeI_BPEH_anyScheme}
The closed-form expression for the OP at $\device^\one_{t+1}$, using the BPEH-powered QoM or CoM scheme, denoted as the PQoM or PCoM scheme, respectively, can be written as
\begin{align}
{\rm OP}^{\rm PQoM}_{\device^\one_{t+1}} &= 
{\rm OP}^{\rm PCoM}_{\device^\one_{t+1}}
\nonumber\\
    &= 1 - 
        \sum_i 
        \sum_j {\rho}^{[i]}_{t+1} {\iota}^{[j]}_{t} 
            \bar{F}_{X_{t+1}}( \tilde{\tau}^{[ij]}_{M} ),
\label{eq_OP_PQoM_PCoM_final}
\end{align}
where
    $\tilde{\tau}^{[00]}_{M} = {\tau}^{[00]}_{M}$,
    $\tilde{\tau}^{[10]}_{M} = {\tau}^{[10]}_{M}$,
    $\tilde{\tau}^{[10]}_{M} \triangleq \frac{ \tau^{[00]}_{M} }{ 1-\beta_{t+1} }$, and
    $\tilde{\tau}^{[11]}_{M} \triangleq \frac{ \tau^{[11]}_{M} }{ 1-\beta_{t+1} }$.
\end{theorem}

\begin{IEEEproof}
Using \eqref{eq_rate_BPEH}, the OP of $\device^\one_t$, in the $t$-th time slot, can be expressed as in \eqref{eq_OP_BTEH_proof_1}, with ${\sf R}^\one_{t+1}$ is given by \eqref{eq_rate_BPEH}. Next, considering the event where there exists at least one $\device^\two$ near $\device^\one_t$, the probability $\Pr[ {\sf R}^\one_{t+1} \ge R_M, \mathbbm{I}_{t+1} = 1 ]$, denoted as ${\cal P}'$, is rewritten as
\begin{align}
&\mathcal{P}'
    = \Pr[ \mathbbm{I}_{t+1} \!=\! 1 ]\!
    \Pr\!\bigg[ 
            (1\!-\!\beta_{t+1}) X_{t+1} \!\ge\! \tau^{[00]}_{M}, 
            \mathbb{M}^{c}_t
    \bigg] \nonumber \\
&\qquad
    + \Pr\bigg[ 
    \frac{p_M (1\!-\!\beta_{t+1}) X_{t+1}}{(1\!-\!p_M) (1\!-\!\beta_{t+1}) X_{t+1}\!+\!1} \!\ge\! \tau^{[00]}_{M}, 
    \mathbb{M}_t
\bigg]
\nonumber\\
&\qquad\qquad\qquad\qquad\qquad\qquad
\times \Pr[ \mathbbm{I}_{t+1} \!=\! 1 ].
\end{align}

Using the identity $\Pr[V]\Pr[U \!\mid\! V] \!=\! \Pr[U V]$ for~two RVs $U$ and $V$, and some mathematical manipulations, the above equation can be further rewritten~as
\begin{align}
\mathcal{P}'
    = \rho_{t+1} 
    [   \iota^{[0]}_t
        \Pr[ X_{t+1} \ge \tilde{\tau}^{[10]}_{M} ] 
        \!+\! \iota^{[1]}_{t}
        \Pr[ X_{t+1} \ge \tilde{\tau}^{[11]}_{M} ]  ].
\end{align}

Invoking \eqref{eq_F_Xm_0} and \eqref{eq_F_Xm_1} in Theorem \ref{theorem_TypeI_BPEH_anyScheme}, we can obtain the CCDF of $X_{t+1}$. In addition, the probability $\Pr[ {\sf R}^\one_{t+1} \ge R_M, \mathbbm{I}_{t+1} = 0 ]$ in \eqref{eq_OP_BTEH_proof_1} can be derived in an analogous manner. 
This completes the proof of Theorem \ref{theorem_TypeI_BPEH_anyScheme}.
\end{IEEEproof}

In \eqref{eq_OP_PQoM_PCoM_final}, the target transmission rate $R_M$ should satisfy
\begin{align}
    R_M < \frac{1}{M-1} \log_2\left( 1 + \frac{p_M}{1-p_M} \right),
\label{eq_target_rate_condition_BPEH}
\end{align}
so that $\bar{F}_{X_{t+1}}( \tilde{\tau}^{[01]}_{M} )$ and $\bar{F}_{X_{t+1}}( \tilde{\tau}^{[11]}_{M} )$ are positive. Accordingly, the MRTR of $s_M$ using the PCoM and PQoM schemes, defined as the right-hand side of \eqref{eq_target_rate_condition_BPEH}, is higher than using TCoM and TQoM schemes.

\subsection{Performance Analysis of Type-II MTCDs}
\label{subsec_PA_typeII_MTCD}

The paired $\device^\two$s require the correct decoding of their own information signals, which depends on the correct decoding of $s_M$. 
    Although each paired $\device^\two$ does not utilize EH, its performance can be affected by whether its own dedicated $\device^\one$ adopts EH. 
It is recalled that both the TCoM and TQoM schemes use the BTEH architecture to power the transmission from $\device^\one$ to $\device^\two$s, where TCoM applies the CoM scheme to schedule more than one $\device^\two$s whereas TQoM only serves an additional $\device^\two$.
    Meanwhile, both the PCoM and PQoM schemes adopt the BPEH architecture for regenerative transmission with a similar use of the CoM and QoM schemes.

Considering that the CoM scheme is adopted at the $\device^\one$s, the OP at each $\device^\two_{t,k}$, given that the dedicated $\device^\one_t$ adopts BTEH with probability $\rho_t$, can be formulated as
\begin{align}
\text{OP}^{\rm TCoM}_{\device^\two_{t,k}}
    \!=\! 1 \!-\! \Pr\left[
    	\big( \mathsf{R}^\two_{t,k} \!\ge\! R_M \big),
    	\!\bigcap_{n = k}^{K_t}\!
    		\big( \mathsf{R}^{\two,[n]}_{t,k} \!\ge\! R_{t,n} \big)
    \right], \label{eq_OP_BTEH_MTCDI_CoM}
\end{align}
which can be derived from the following Theorem.
\begin{theorem} \label{theorem_TypeII_BTEH_COM}
The exact closed-form expression of the OP at $\device^\two_{t,k}$, using the TCoM scheme, is given by 
\begin{align}
{\rm OP}^{{\rm TCoM}}_{\device^\two_{t,k}}
	&= 1 - \sum_{i} \rho_{t+1}^{[i]} \bar{F}_{Y_{t,k}}( \tau^{[i1]}_{t,k}  ),
\label{eq:OP:UE:BTEH:QoM}
\end{align}
where for
    $\tau^{[00]}_{t,k} \triangleq 2^{(M-1) R_{t,k}}-1$ and 
	$\tau^{[10]}_{t,k} \triangleq 2^{\frac{M-1}{1-\alpha_{t+1}} R_{t,k}}-1$, we have
\begin{align}
    \tau^{[i1]}_{t,k} 
    &\triangleq \max\limits_{\mathclap{k \le n \le K_t}}\,\,\left[ \tau^{[i0]}_{M}, \frac{ \tau^{[i0]}_{t,k} }{ p_{t,n} - \tau^{[i0]}_{t,k} \sum_{q=1}^{n-1}{p_{t,q}} } \right], \nonumber 
\end{align}
\end{theorem}

\begin{IEEEproof}
Substituting \eqref{eq_gamma_nq_m1_nk_end} and \eqref{eq_gamma_m1_nk_end} into \eqref{eq_OP_BTEH_MTCDI_CoM} with a note that $\frac{P_t \varphi_{t,k}}{\sigma^2} = Y_{t,k}$, we obtain
\begin{align}
&{\rm OP}^{{\rm TCoM}}_{\device^\two_{t,k}}
    = 1 
    - \sum_{i}
    \Pr\bigg[
        \frac{p_M Y_{t,k}}{(1-p_M) Y_{t,k}+1} \ge \tau^{[i0]}_{M},
\nonumber\\
    &\quad
        \bigcap_{n=k}^{K_t}
            \frac{p_{t,n} Y_{t,k}}{Y_{t,k}\sum_{q=1}^{n-1}{ p_{t,q} }+1} \ge \tau^{[i0]}_{t,k},
            \big[ 
                \mathbbm{I}_{t+1} = i
            \big] 
    \bigg].\!\!
\label{eq_OP_BTEH_CoM_1}
\end{align}

After some mathematical manipulations, the above equation can be rewritten as
\begin{align}
{\rm OP}^{{\rm TCoM}}_{\device^\two_{t,k}}
    = 1 - \sum_{i} 
    \rho^{[i]}_{t+1}
    \Pr[ Y_{t,k} \ge \tau^{[i1]}_{t,k} ].
\label{eq_OP_BTEH_CoM_2}
\end{align}

The CDF of the received power normalized by the noise power at each paired ${\rm D}_{t,k}^{\rm [II]}$ using the CoM scheme is given by 
\begin{align}
&\bar{F}_{Y_{t,k}}(y)
   	= {\rho}_t^{[0]}
    	\bar{F}_{\varphi_{t,k}}\left( \frac{y}{ \bar{\gamma}_0 } \right) 
    + \frac{2}{\epsilon}
    \sum_{\tau=1}^{t-1}
    \bigg[ 
        {{\rho}_{\tau}^{[0]} \prod_{j=\tau+1}^{t} \rho_j^{[1]}}
    \bigg]
\nonumber\\
&\times
    \Bigg\{
    \chi^{[0]}_{k}
    G^{t-\tau+1,1}_{1,t-\tau+2}
	\left[
		\frac{ 
		\left(
		    \frac{k}{K_t}
		\right)^\epsilon }{ \bar{\gamma}_0 \ell(r_{t}) }
		\frac{y}{ \bar{\xi}_{\tau+1,t}  }
		\left\vert 
		\begin{array}{cc}
		   	1 - \frac{2}{\epsilon} \\
		   	{\bf 1}_{t-\tau}, 0,-\frac{2}{\epsilon} 
		\end{array}
		\right.
	\right]  \nonumber\\
	&\quad- \left. \chi^{[1]}_{k}
    G^{t-\tau+1,1}_{1,t-\tau+2}
	\left[
		\frac{ \left(
		    \frac{k-1}{K_t}
		\right)^\epsilon }{ \bar{\gamma}_0 \ell(r_{t}) }
		\frac{y}{ \bar{\xi}_{\tau+1,t}  }
		\left\vert 
		\begin{array}{cc}
		   	1 - \frac{2}{\epsilon} \\
		   	{\bf 1}_{t-\tau}, 0,-\frac{2}{\epsilon} 
		\end{array}
		\right. \!\!
	\right] \! \right\}
\nonumber\\
&\qquad\qquad\qquad\qquad\qquad\qquad\qquad\quad,~y > 0.\!\!\label{eq:CCDF_Y_tk}
\end{align}
The derivation of \eqref{eq:CCDF_Y_tk} is provided in Appendix \ref{apendix_proof_lemma_TypeII_CoM_Ytk}.

Substituting \eqref{eq:CCDF_Y_tk} into \eqref{eq_OP_BTEH_CoM_2}, we obtain \eqref{eq:OP:UE:BTEH:QoM}. This completes the proof of Theorem \ref{theorem_TypeII_BTEH_COM}.
\end{IEEEproof}
It is worth noting that $\bar{F}_{Y_{t,k}}(y)$ is only defined for $y > 0$, thus $\bar{F}_{Y_{t,k}}( \tau^{[01]}_{t,k} ) = 0$ and $\bar{F}_{Y_{t,k}}( \tau^{[11]}_{t,k} ) = 0$ when $\tau^{[01]}_{t,k} < 0$ and $\tau^{[11]}_{t,k} < 0$, or, equivalently,
\begin{align}
    \frac{ \tau^{[i0]}_{t,k} }{ p_{t,n}-\tau^{[i0]}_{t,k} \sum_{q=1}^{n-1}{p_{t,q}} < 0 }.
\label{eq_44}
\end{align}
This also implies that the target transmission rates of every $s_{t,n}$ so that $\bar{F}_{Y_{t,k}}( \tau^{[01]}_{t,k} )$ and $\bar{F}_{Y_{t,k}}( \tau^{[11]}_{t,k} )$ are theoretically positive should satisfy
\begin{align}
R_{t,n} < \frac{1-\alpha_{t+1}}{M-1} 
    \log_2\left(
        1 + \frac{ p_{t,n} }
            { \sum_{q=1}^{n-1}{p_{t,q}} }
    \right),
\label{eq_TCoM_MRTR}
\end{align}
which is obtained by solving \eqref{eq_44} for $R_{t,n}$.
    The right-hand side of \eqref{eq_TCoM_MRTR} is the MRTR of $s_{t,n}$ using the TCoM scheme. 
It is noted that the MRTR of the $\device^\two_{t,1}$'s information signal is undefined since the intra-interference power while decoding $s_{t,1}$ is perfectly cancelled, i.e., $\sum_{q=1}^{n-1}{p_{t,q}} = 0$ when $n = 1$.

Using the BPEH architecture, the decoding threshold of $s_{t,k}$ is not affected by whether $\device^\one_{t+1}$ adopts EH. 
Specifically, the outage probability of $\device^\two_{t,k}$ can be formulated analogously to \eqref{eq_OP_BTEH_MTCDI_CoM}, in which $\tau_{t+1} = \frac{1}{M-1}$,
which is derived in the following Theorem.

\begin{theorem} \label{theorem_TypeII_BPEH_COM}
The exact closed-form expression of the OP at $\device^\two_{t,k}$, using the PCoM scheme, is given by
\begin{align}
{\rm OP}^{\rm PCoM}_{\device^\two_{t,k}} 
    =   1 - \bar{F}_{Y_{t,k}}( \tau^{[01]}_{t,k} ).
\label{eq_OP_PCoM_fin}
\end{align}
\end{theorem}

\begin{IEEEproof}
From \eqref{eq_OP_BTEH_CoM_1} and \eqref{eq_OP_BTEH_CoM_2}, the OP at $\device^\two_{t,k}$ for the PCoM scheme can be derived as
\begin{align}
{\rm OP}^{\rm PCoM}_{\device^\two_{t,k}}
&= 1 - \rho_{t+1}^{[0]}
    \Pr[ Y_{t,k} \ge \tau^{[01]}_{t,k} ]
\nonumber\\
&\qquad
    - \rho_{t+1}^{[1]}
    \Pr[ Y_{t,k} \ge \tau^{[01]}_{t,k} ].
\label{eq_OP_BTEH_CoM_3}
\end{align}
Invoking \eqref{eq:CCDF_Y_tk} in Theorem \ref{theorem_TypeII_BTEH_COM} to obtain the CCDF of $Y_{t,k}$, we rewrite the above equation as in \eqref{eq_OP_PCoM_fin}. This completes the proof of Theorem \ref{theorem_TypeII_BPEH_COM}.
\end{IEEEproof}

In the BTEH-powered QoM scheme, the outage event is avoided by correctly decoding both $s_M$ and the information signal of $\device^\two_{t,k^\ast}$, i.e., $s_{t,k^\ast}$. 
    Thus, the OP of the paired $\device^\two_{t,k^\ast}$ during the $t$-th time slot can be expressed as
\begin{align}
{\rm OP}^{\rm TQoM}_{\device^\two_{t,k^\ast}}
	\!=\! 1 \!-\! \Pr[ \mathsf{R}^{\two}_{M \to t} \ge R_M, \mathsf{R}^{\two}_{t} \ge R_{t,k^\ast} ],
\label{eq_OP_BTEH_YoM}
\end{align}
which is derived in the following Theorem.

\begin{theorem} \label{theorem_TypeII_BTEH_QOM}
The exact closed-form expression of the OP at $\device^\two_{t,k^\ast}$, using the TQoM scheme, is given by
\begin{align} 
{\rm OP}^{\rm TQoM}_{\device^\two_{t,k^\ast}}
	= 1 - \sum_i \rho^{[i]}_{t+1} \bar{F}_{Z_{t}}( \tau^{[i1]}_{t,k^\ast} ), 
\label{eq_OP_BTEH_QoM_final}
\end{align} 
where for $\scriptstyle {\tau^{[00]}_{t,k^\ast} \triangleq 2^{(M - 1) R_{t,k^\ast}} \!\!\!-\! 1}$ and $\scriptstyle {\tau^{[10]}_{t,k^\ast} \triangleq 2^{\frac{M\!-\!1}{1 \!-\! \alpha_{t+1}} R_{t,k^\ast}} - 1}$, we have 
    ${\tau^{[i1]}_{t,k^\ast} \triangleq
	\max\!\left[
		\tau^{[i1]}_{M} , \frac{ \tau^{[i0]}_{t,k^\ast} }{ 1 - p_M }
	\!\right]}$.
\end{theorem}

\begin{IEEEproof}
Substituting \eqref{eq_RM2t} and \eqref{eq_Rt} into \eqref{eq_OP_BTEH_YoM}, and after some mathematical simplifications, we obtain
\begin{align}
{\rm OP}^{\rm TQoM}_{\device^\two_{t,k^\ast}}
    &= 1 - 
    \sum_i
    \Pr\bigg[
        \frac{p_M Z_t}{(1-p_M) Z_t+1} \ge \tau^{[i0]}_{M}, 
\nonumber\\
    &\qquad~
        (1-p_M) Z_t \ge \tau^{[i0]}_{t,k^\ast},
        \mathbbm{I}_{t+1} = i
    \bigg].\!\!
\end{align}

After some mathematical manipulations, the above equation can be rewritten as
\begin{align}
{\rm OP}^{\rm TQoM}_{\device^\two_{t,k^\ast}}
    &= 1 - 
    \sum_i \rho^{[i]}_{t+1} 
    \Pr[ Z_t \ge \tau^{[i1]}_{t,k^\ast} ].
\label{eq_OP_BTEH_QoM_proof_2}
\end{align}
which is equivalent to \eqref{eq_OP_BTEH_QoM_final} as $\Pr[ Z_t \ge z ] = \bar{F}_{Z_t}(z)$, i.e., the CCDF of $Z_{t}$.
    Recalling that $Z_t \triangleq P_t \varphi_t/\sigma^2$, thus it is required to obtain the distribution of $\varphi_t$ before deriving the CCDF of $Z_t$.
Accordingly, the CDF of the received power observed at each $\device^\two_{t,k^\ast}$ during the $t$-th time slot for the QoM scheme can be expressed as
\begin{align}
F_{\varphi_t}(\varphi)
	= 1 - \frac{1}{\Gamma(m_t)}
	H^{1,1}_{1,1}
	\bigg[
	    \bigg( 
		    \frac{\varphi}{\mu_t}
		\bigg)^{-\theta_t}
		\bigg\vert
		\begin{array}{cc}
			 (1,1) \vspace{1pt} \\ (m_t,1)
		\end{array}
	\bigg],
\label{eq_CDF_varphi_t}
\end{align}
where $\varphi > 0$, $H^{m,n}_{p,q}[\cdot]$ is the Fox H-function \cite{Kilbas2004}.
\begin{IEEEproof}
The proof is provided in Appendix \ref{apx:B}.
\end{IEEEproof}

The CCDF of the received power normalized by the Gaussian noise at each paired $\device^\two$ in the QoS schemes is expressed as
\begin{align}
&\bar{F}_{Z_t}(z)
    = {\rho}^{[0]}_t
	\bar{F}_{\varphi_t}
	\left(
		\frac{z}{\bar{\gamma}_0}
	\right)	+ \sum_{\tau=1}^{t-1}
    	\frac{ {\rho}^{[0]}_{\tau} \prod_{{j=\tau+1}}^{t} \rho^{[1]}_j }{\Gamma(m_t)} 
\nonumber\\
    &\times
\scriptstyle
    H^{t-\tau+1,1}_{1,t-\tau+1}
    \Bigg[
    	\bigg(
    		\frac{z}{\bar{\xi}_{\tau+1,t} \mu_t \bar{\gamma}_0}
    	\bigg)^{\theta_t}
    	\bigg\vert\!\!\!
    	\begin{array}{cc}
    		(1-m_t,1) \vspace{0.5pt}\\
    		(0,1), \underbrace{(1,\theta_t),\cdots,(1,\theta_t)}\limits_{t-\tau~{\rm terms}}
    	\end{array}\!
    \Bigg],
\label{eq_F_Z_t}
\end{align}
for $z > 0$.
\begin{IEEEproof}
The proof is provided in Appendix \ref{apendix_proof_lemma_Z_fin}.
\end{IEEEproof}

Substituting the above equation into \eqref{eq_OP_BTEH_QoM_proof_2}, we obtain \eqref{eq_OP_BTEH_QoM_final}. This completes the proof of Theorem~ \ref{theorem_TypeII_BTEH_QOM}.
\end{IEEEproof}

In \eqref{eq_OP_BTEH_QoM_final}, the target transmission rate of $R_{t,k^\ast}$ should satisfy $\tau^{[01]}_{t,k^\ast} > 0$ and $\tau^{[11]}_{t,k^\ast} > 0$ otherwise the corresponding OP becomes one.

\begin{theorem} \label{theorem_TypeII_BPEH_QOM}
The exact closed-form expression of the OP at $\device^\two_{t,k^\ast}$, using the PQoM scheme, is given by
\begin{align}
{\rm OP}^{\rm PQoM}_{\device^\two_{t,k^\ast}} 
    = 1 - \bar{F}_{Z_t}( \tau^{[01]}_{t,k^\ast} ). \label{eq_OP_BPEH_QoM_final}  
\end{align}
\end{theorem}

\begin{IEEEproof}
From \eqref{eq_RM2t} and \eqref{eq_Rt}, the outage probability of $\device^\two_{t,k^\ast}$ can be expressed as
\begin{align}
&{\rm OP}^{\rm PQoM}_{\device^\two_{t,k^\ast}} 
    = 1 - \Pr\bigg[
        \frac{p_M Z_t}{(1-p_M) Z_t+1} \ge \tau^{[00]}_{M},
\nonumber\\
    &\quad
    (1-p_M) Z_t \ge \tau^{[00]}_{t,k^\ast}
    \bigg]
    = 1 - \Pr[  Z_t \ge \tau^{[01]}_{t,k^\ast} ].
\end{align}

By invoking \eqref{eq_F_Z_t}, we obtain \eqref{eq_OP_BPEH_QoM_final}. This completes the proof of Theorem \ref{theorem_TypeII_BPEH_QOM}.
\end{IEEEproof}

From \eqref{eq_OP_BPEH_QoM_final} and \eqref{eq_OP_BTEH_QoM_final}, the MRTR of $\device^\two_{t,k^\ast}$'s information signal in the TQoM and PQoM schemes is undefined since the intra-interference power while decoding $s_{t,k^\ast}$ is perfectly canceled.

\subsection{Sum-Throughput Analysis of the Proposed mMTC Network}
\label{subsec_SumT_mMTC}

In this subsection, the e2e OPs of the MTCDs and the network sum-throughput are studied. First, the network sum-throughput is obtained in the following Theorem. 

\begin{theorem} \label{theorem_e2e_throughput_SC}
The closed-form expressions for the sum-throughput can be expressed as
\begin{align}
{\cal T}_{\Sigma} 
	&=  \frac{ R_M }{M-1}
	( 1 - {\rm OP}^{\one,{\rm e2e}}_{M} ) 
\nonumber\\
	&\quad+ \sum_{t=1}^{M-1}
	\sum_{\kappa \in \mathcal{K}_{t} }
	\frac{ R_{t,\kappa} }{M-1} 
	( 1 - {\rm OP}^{\two,{\rm e2e}}_{t,\kappa} ), 
\label{eq_sum_throughput_SC}
\end{align}
where ${\cal K}_{t}$ is the set of $\device^\two$s that are scheduled to perform NOMA. Herein, ${\cal K}_{t} \triangleq \{ 1,2,\dots, K_t \}$ for TCoM and PCoM schemes, and ${\cal K}_{t} \triangleq \{ \kappa: \kappa = k^\ast \}$ for TQoM and PQoM schemes.
Note that the factor $\frac{1}{M-1}$ accounts for the $(M-1)$ time slots consumed 
    in the transmission from $\device^\one_1$ to $\device^\one_M$.
\end{theorem}

\begin{IEEEproof}
In order to characterize the outage performance of the proposed network, it is important to obtain and inspect the e2e OP of $\device^\one_M$. 
    In the proposed schemes, the outage at $\device^\one_M$ depends on whether an outage occurs at other $\device^\one_t$s, $1 \le t \le M$. 
In other words, the e2e OP at $\device^\one_M$ can be expressed~as
\begin{align}
{\rm OP}^{\one,{\rm e2e}}_{M}
    = \Pr[ \bigcup_{t=1}^{M-1} ( \mathsf{R}^{\one}_{t} < R_M ) ], \label{eq_e2eOP_SC_BXEH} 
\end{align}

However, it is intricate and even impossible to obtain the exact analytical expression for the above equation. 
    In the low probability of EH (LPEH) regime, the e2e OP at $\device^\one_{M}$ using scheme ${\sf S}$ can be given by
\begin{align}
{\rm OP}^{\one,{\rm e2e}}_{M}
	\cong 1 - \prod_{t=1}^{M-1} 
	[ 1 - {\rm OP}^{\sf S}_{\device^\one_{t+1}} ],
\label{eq_e2eOP_SC_BXEH_LPEH} 
\end{align}
where ${\sf S} \in \{ {\rm TQoM}, {\rm TCoM}, {\rm PQoM}, {\rm PCoM} \}$.

In addition, the performance of $\device^\two_t$ relies on whether the previous $\device^\one_{\tau}$, $\forall \tau \in [1,t-1]$, and its dedicated transmitter $\device^\one_{t}$ can operate without outage.       
Hence, the e2e OP at the paired $\device^\two$ in TCoM and PCoM schemes can be formulated as
\begin{align}
&
{\rm OP}^{\two,{\rm e2e}}_{t,k}
	= 1 - \Pr[
	    ( \mathsf{R}^{\two}_{t,k} \ge R_M ), 
\nonumber\\
    &\qquad\quad
	\bigcap\limits_{n = k}^{K_t}
	    ( \mathsf{R}^{\two, [n]}_{t,k} \ge R_{t,n} ),
    \bigcap\limits_{i=1}^{t}
	   	( \mathsf{R}^{\one}_{i} \ge R_M )
	], \label{eq_OP_e2e_TCMS_sim}
\end{align}
and the e2e OP at the paired $\device^\two$ in TQoM and PQoM schemes can be formulated as
\begin{align}
{\rm OP}^{\two,{\rm e2e}}_{t,k^\ast}
	&= 1 - \Pr[
	   	( \mathsf{R}^{\two}_{M \to t} \ge R_M ), 
\nonumber\\&
    \bigcap\limits_{i=1}^{t}
   		( \mathsf{R}^{\two}_{t,k^\ast} \ge R_{t,k^\ast} ), 
	   	( \mathsf{R}^{\one}_{i} \ge R_M )
	] \label{eq_OP_e2e_TQMS_sim}.
\end{align}

In the LPEH regime, the above e2e OPs can be asymptotically expressed as
\begin{align}
{\rm OP}^{\two,{\rm e2e}}_{t,\kappa}
	\!\cong\!
	1 \!-\! [ 1 \!-\! {\rm OP}^{\sf S}_{\device^\two_{t,\kappa}} ]
	\prod_{i=1}^t [ 1 \!-\! {\rm OP}^{\sf S}_{\device^\one_{t+1}} ].
\label{eq_e2eOP_DII}
\end{align}
The throughput of the signals $s_M$ and $s_{t,k}$ are defined via the corresponding e2e OPs. 
    Adding the sum of \eqref{eq_e2eOP_DII} over $t \in [1,M-1]$ and $k \in {\cal K}_{t}$ we obtain \eqref{eq_sum_throughput_SC}. 
This completes the proof of Theorem~\ref{theorem_e2e_throughput_SC}.
\end{IEEEproof}

\section{Asymptotic Performance Analysis}

In this section, we provide approximations for the e2e OP, and consequently the sum-throughput of the proposed schemes.

\subsection{Asymptotic OP of $\device^\one$}

\begin{lemma}
\label{lem_asymp_X}
In the high transmission power regime, $\bar{\gamma}_{0} \to \infty$, $F_{X_{t+1}}(x)$ can be approximated as
\begin{align}
F_{X_{t+1}}(x) 
    \to 
    F_t\left( \frac{x}{\bar{\gamma}_0 \ell(l_{t+1})} \right) 
    \frac{x}{\bar{\gamma}_0 \ell(l_{t+1})},
\label{eq_X_asymp}
\end{align}
where
\begin{align}
{{F}_{t}}(x)
    &\triangleq 
    \rho^{[0]}_{t} + \sum_{\tau=1}^{t-1} \frac{ \rho^{[0]}_{\tau} \prod_{j=\tau+1}^{t} \rho^{[1]}_{j} }{ \bar{\xi}_{\tau+1,t} }
    \sum_{n=0}^{t-\tau} \frac{\psi^{(n)}_{t-\tau+1}}{n!}
\nonumber\\&\qquad\quad\times
    \sum_{r=0}^{t-\tau-n} \frac{(-1)^r}{r!}
    \bigg(
        \log \frac{x}{ \bar{\xi}_{\tau+1,t} } 
    \bigg)^r.
\end{align}
\end{lemma}
\begin{IEEEproof}
The proof is provided in Appendix \ref{apx_proof_F}.
\end{IEEEproof}

\subsection{The Asymptotic OP at $\device^\two$}

Given that $\gamma(a;x) \to \frac{x^a}{a} - \frac{x^{a+1}}{a+1}$ and $e^{-x} \to 1-x$ as $x \to 0$, we have
\begin{align}
F_{\varphi_{t,k}}(\varphi)
    \to \bar{\varphi}_{t,k}
    \frac{\varphi}{\ell(r_t)},
~
F_{\varphi_t}(\varphi)
    \to \bar{\varphi}_{t}
    \frac{\varphi}{\ell(r_t)},~\varphi \to 0,
\end{align}
where
$
\bar{\varphi}_{t,k} 
\triangleq
\frac{ \frac{2}{\epsilon} }{ \frac{2}{\epsilon}+1 }
\big[ 
    {\chi}^{[0]}_{k} \big( \frac{k}{K_t} \big)^{\epsilon}
    -   {\chi}^{[1]}_{k} \big( \frac{k-1}{K_t} \big)^{\epsilon}
\big]
$ and
$\bar{\varphi}_{t} 
\triangleq \frac{ (\pi \lambda_t r_t^2)^{-\frac{\epsilon}{2}} }{ 1-e^{-\pi \lambda_t r_t^2} }
\gamma\big( \frac{\epsilon}{2}+1; \pi \lambda_t r_t^2 \big)
$.
\begin{lemma}
\label{lem_asymp_YZ}
In the high transmission power regime, ${F_{Y_{t,k}}(y)}$ and ${F_{Z_t}(z)}$ can be approximated as
\begin{align}
F_{Y_{t,k}}(y) 
    &\to F_t\bigg( \frac{\bar{\varphi}_{t,k} y}{\bar{\gamma}_0 \ell(r_t)} \bigg)
    \frac{ \bar{\varphi}_{t,k} }{\bar{\gamma}_0 \ell(r_t)} y,~
\bar{\gamma}_{0} \to \infty, \label{eq_Y_asymp} \\
F_{Z_t}(z) 
    &\to F_t\bigg( \frac{\bar{\varphi}_{t} z}{\bar{\gamma}_0 \ell(r_t)} \bigg)
    \frac{ \bar{\varphi}_{t}}{\bar{\gamma}_0 \ell(r_t)} z,~
\bar{\gamma}_{0} \to \infty. \label{eq_Z_asymp} 
\end{align}
\end{lemma}
\begin{IEEEproof}
In Lemma \ref{lem_asymp_YZ}, we obtain \eqref{eq_Y_asymp} and \eqref{eq_Z_asymp} by adopting mathematical steps shown in Appendix~\ref{apx_proof_F}. 
Hence, we omit them from the paper.
\end{IEEEproof}

\begin{table*}[!h]
	\centering
	\caption{Simulation Parameters}{
		\begin{tabularx}{\linewidth}{|X|l|X|l|}
			\Xhline{2\arrayrulewidth}
			\textbf{Parameter} & \textbf{Value} & \textbf{Parameter} & \textbf{Value} \\
			\Xhline{2\arrayrulewidth}
			{Number of $\device^\one$, $M$, [device]} & {$4$} & {Active $\device^\two$ density, $\lambda^{[1]}_m$ [${\rm device/m}^2$]} & {$1$e-$2$} \\
			{Antenna gains of $\device^\one$, [dBi]} & {$5$ \cite{Do_TCOM_2018}} & {Inactive $\device^\two$ density, $\lambda^{[0]}_m$ [${\rm device/m}^2$]} & {$1$e-$3$} \\
			{Antenna gains of $\device^\two$, [dBi]} & {$5$ \cite{Do_TCOM_2018}} & {Carrier frequency, $f_c$ [GHz]} & {3 \cite{DoTCOM2021}} \\
			{Bandwidth, $BW$ [MHz]} & {$10$ \cite{Do_TCOM_2018}} & {Fixed transmission power, $P_0$ [dBm]} & {$0$} \\
			{Noise power density, $\sigma^2$ [dBm/Hz]} & {-$174$ \cite{FengWCL2021}} & {Energy conversion efficiency, $\eta_m$} & {$1$ \cite{AtatTCOMM2017}} 
			\\
			\Xhline{2\arrayrulewidth}
	\end{tabularx} }
	\label{table_parameters}
\end{table*}

\subsection{Diversity order}

Using the asymptotic results of $F_{X_{t+1}}(x)$ in Lemma \ref{lem_asymp_X}, $F_{Y_{t,k}}(y)$, and $F_{Z_{t}}(z)$ in Lemma \ref{lem_asymp_YZ}, we find that the ${\rm OP}^{\one,\rm e2e}_{M}$, ${\rm OP}^{\two, \rm e2e}_{t,k}$, and ${\rm OP}^{\two, \rm e2e}_{t,k^\ast}$ have no error floors, respectively, when increasing $\bar{\gamma}_{0}$ (i.e., increasing $P_0$).
    In this context, the diversity order which is important to evaluate the system performance in the high transmission regime, is defined as \cite{Do_TCOM_2018}
\begin{align}
{\sf D}^{\one}_{M}
    &= - \lim\limits_{\bar{\gamma}_{0} \to \infty} 
    \frac{ \log_r {\rm OP}^{\one, \rm e2e}_{M} }
        { \log_r \bar{\gamma}_{0} }, 
\nonumber\\
{\sf D}^{\two}_{t,k}
    &= - \lim\limits_{\bar{\gamma}_{0} \to \infty} 
    \frac{ \log_r {\rm OP}^{\two, \rm e2e}_{t,k} }
        { \log_r \bar{\gamma}_{0} },
\nonumber\\
{\sf D}^{\two}_{t,k^\ast}
    &= - \lim\limits_{\bar{\gamma}_{0} \to \infty} 
    \frac{ \log_r {\rm OP}^{\two, \rm e2e}_{t,k^\ast} }
        { \log_r \bar{\gamma}_{0} },
\end{align}
in which the coefficient $r > 0$ does not affect the diversity order.
It is noted that when $\bar{\gamma}_{0}$ is relatively large, increasing $\bar{\gamma}_{0}$ by $r$ [dB] results in a decrease in ${\rm OP}^{\one, \rm e2e}_{M}$, ${\rm OP}^{\two, \rm e2e}_{t,k}$, and ${\rm OP}^{\two, \rm e2e}_{t,k^\ast}$ by ${\sf D}^{\one}_{M} \cdot r$, ${\sf D}^{\two}_{t,k} \cdot r$, and ${\sf D}^{\two}_{t,k^\ast} \cdot r$ [dB], respectively.
    In the proposed system, with ${F}_{X_{t+1}}(x)$, ${F}_{Y,{t,k}}(y)$, and $F_{Z_{t}}(z)$ given in \eqref{eq_X_asymp}, \eqref{eq_Y_asymp}, and \eqref{eq_Z_asymp}, respectively, 
    we obtain 
$
    {\sf D}^{\one}_{M} = {\sf D}^{\two}_{t,k} = {\sf D}^{\two}_{t,k^\ast} = 1. 
$
This result implies that the e2e OPs at $\device^\one_M$ and $\device^\two_{t,k}$ for the CoM scheme, and $\device^\two_{t,k^\ast}$ for the QoM scheme can be reduced by $x$ times when $\bar{\gamma}_{0}$ increases by $x$ times in the high transmission power regime. In addition, when the diversity order is non-zero, the sum-throughput of mMTC-NOMA can be approximated as
$
{\cal T}_{\Sigma} 
	\to  
	\frac{ 1 }{M-1}
	(
    	R_M + \sum_{t=1}^{M-1} \sum_{\kappa \in \mathcal{K}_{t} } R_{t,\kappa}
	)$.

\section{Numerical Results}

\begin{figure*}[t]
\centering
\setlength{\tabcolsep}{0pt}
\begin{tabular}{cccc}
    \includegraphics[width=0.35\linewidth]{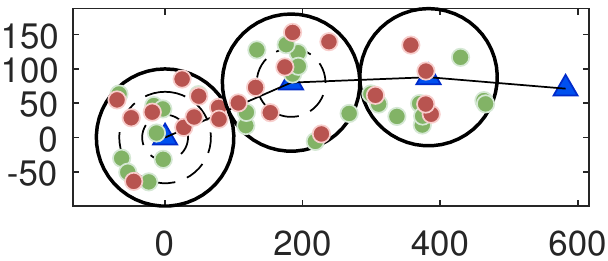} & \multicolumn{2}{c}{ \includegraphics[width=0.35\linewidth]{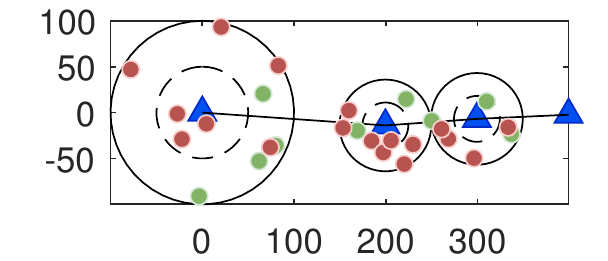}} &
    \includegraphics[width=0.35\linewidth]{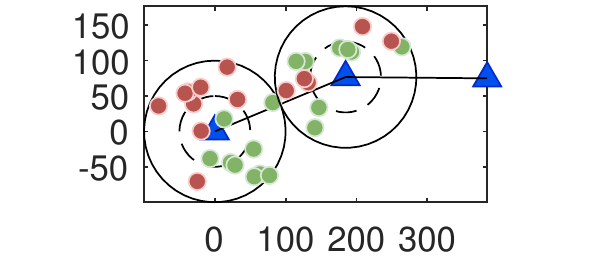} 
    \\
    { $\scriptstyle \text{(a) } {\bf T}_1$.} & \multicolumn{2}{c}{$ \text{(b) } \scriptstyle {\bf T}_2$.} & {$\text{(c) } \scriptstyle  {\bf T}_3$.}
    \\
    \multicolumn{4}{c}{
    \includegraphics[width=0.35\linewidth]{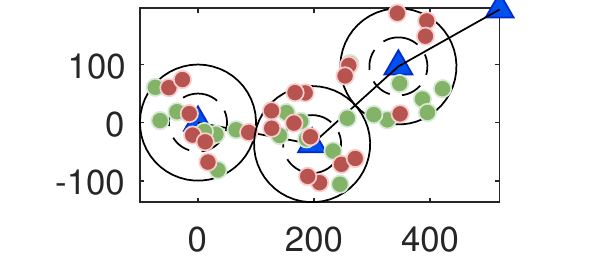}
    \quad
    \includegraphics[width=0.35\linewidth]{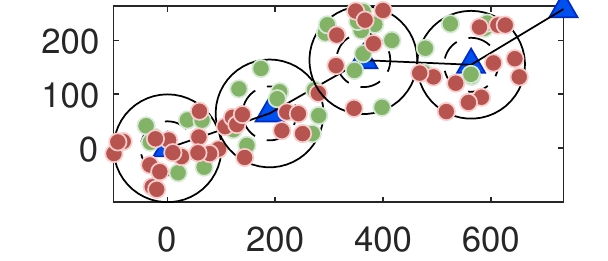}} 
    \\
    \multicolumn{4}{c}{$\text{(d) } \scriptstyle  {\bf T}_4$. \hspace{0.25\linewidth} $\text{(e) } \scriptstyle {\bf T}_5$.}
    \\
\end{tabular}
\caption{ Network topology, where ${\bf T}_i = \left(\begin{smallmatrix}
    l_1 & l_2 & \cdots & l_{M-1} \\
    r_1 & r_1 & \cdots & r_{M-1} \\
    K_1 & K_2 & \cdots & K_{M-1}
\end{smallmatrix}\right)$. The solid blue triangles are $\device^\one$s, active and inactive $\device^\two$s are the solid green and red circles, respectively.
\label{fig_network_topo}}
\end{figure*}

\begin{figure}
	\centering
	\adjincludegraphics[width=\linewidth,valign=T]{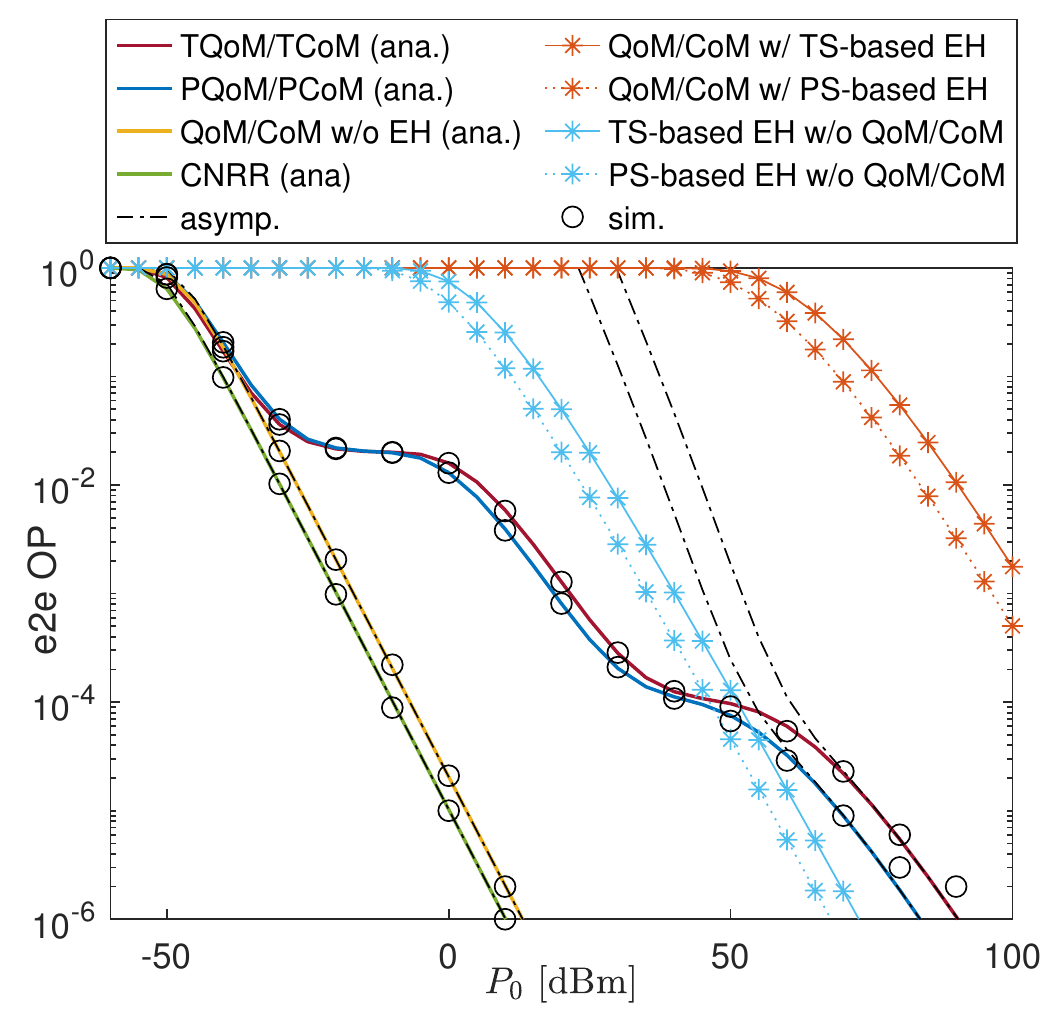} 
	\caption{ The e2e outage probability of $\device^\one_{M}$ as a function of the fixed transmission power, $P_0$, [dBm].}
	\label{fig_e2e_OP_vs_P0}
\end{figure}

\begin{figure}
	\centering
	\adjincludegraphics[width=\linewidth,valign=T]{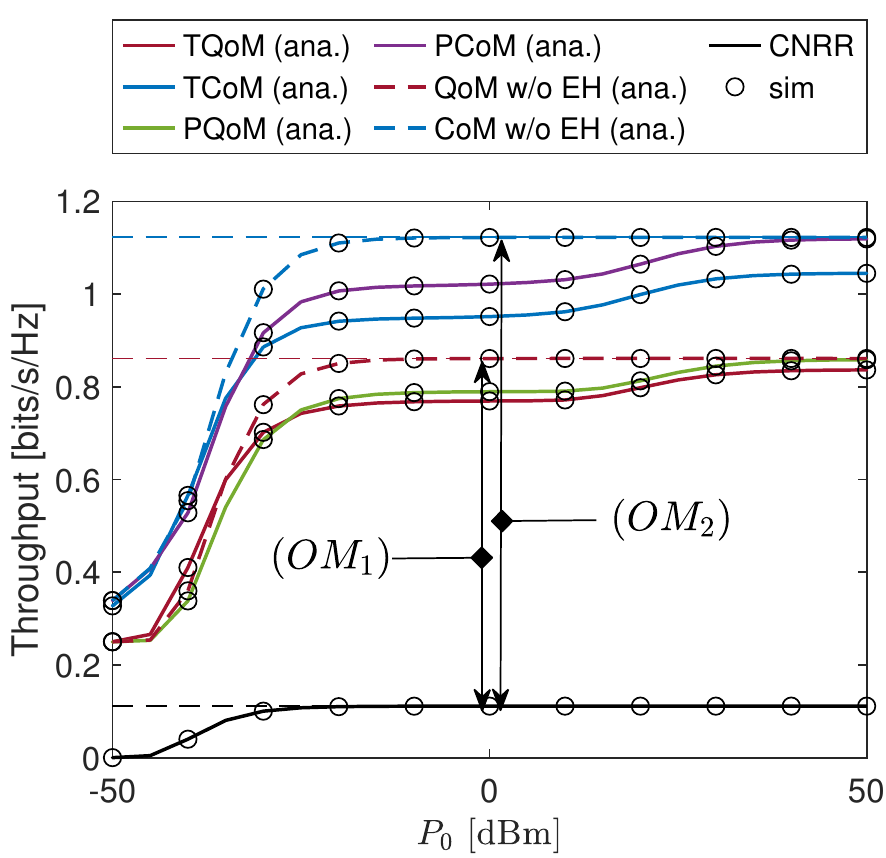}
	\caption{ Network sum-throughput [bits/s/Hz] versus the fixed transmission power, $P_0$ [dBm].}
	\label{fig_throughput_vs_P0}
\end{figure}

\begin{figure}[!h]
    \centering
    \includegraphics[width = \linewidth]{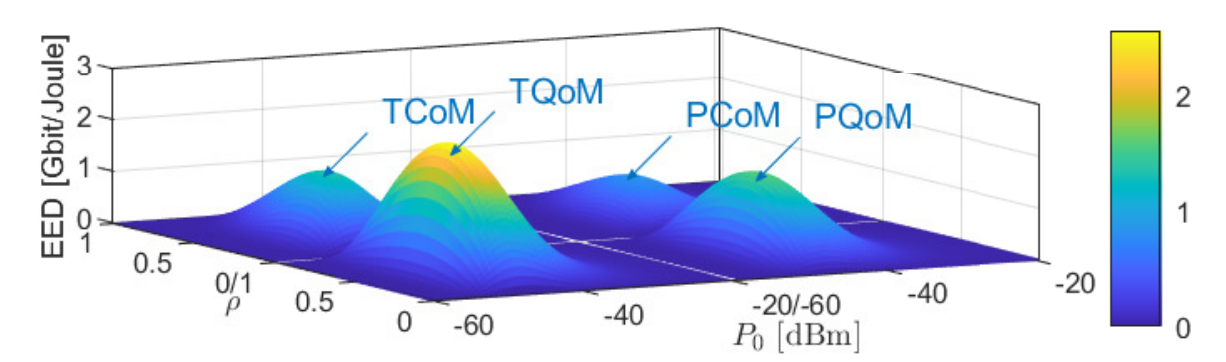}
    \caption{ Energy Efficiency Difference (EED) among proposed transmission schemes, namely TCoM, TQoM, PCoM, and PQoM, as a function of $\rho$ and $P_0$ [dBm].}
    \label{fig_COM_1p4_2}
\end{figure}

\begin{figure}[!h]
    \centering
	\adjincludegraphics[width=\linewidth,valign=T]{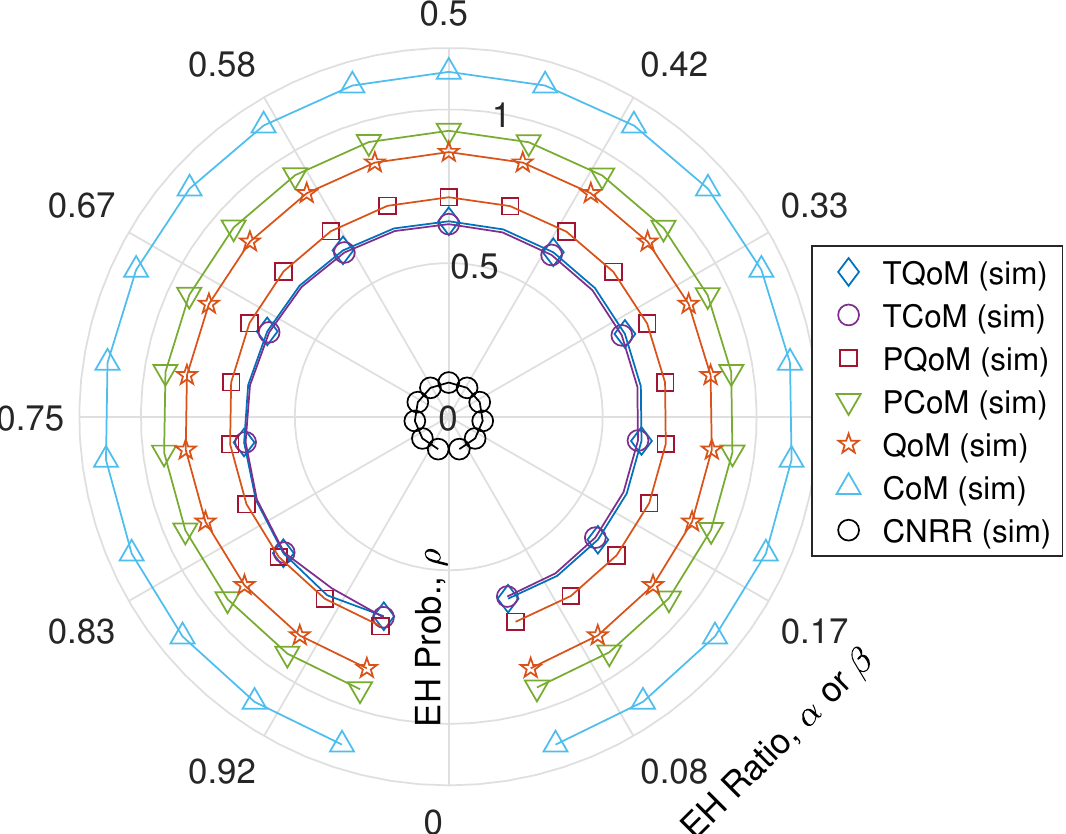} 
	\caption{ Throughput [bits/s/Hz] versus EH ratios (i.e., $\alpha$ or $\beta$), and probability of EH (i.e., $\rho$). Solid lines denote analytical results.}
	\label{fig_throughput_2D_vs_EH_Ratio}
\end{figure}

\begin{figure}[!h]
	\centering
	\adjincludegraphics[width=\linewidth,valign=T]{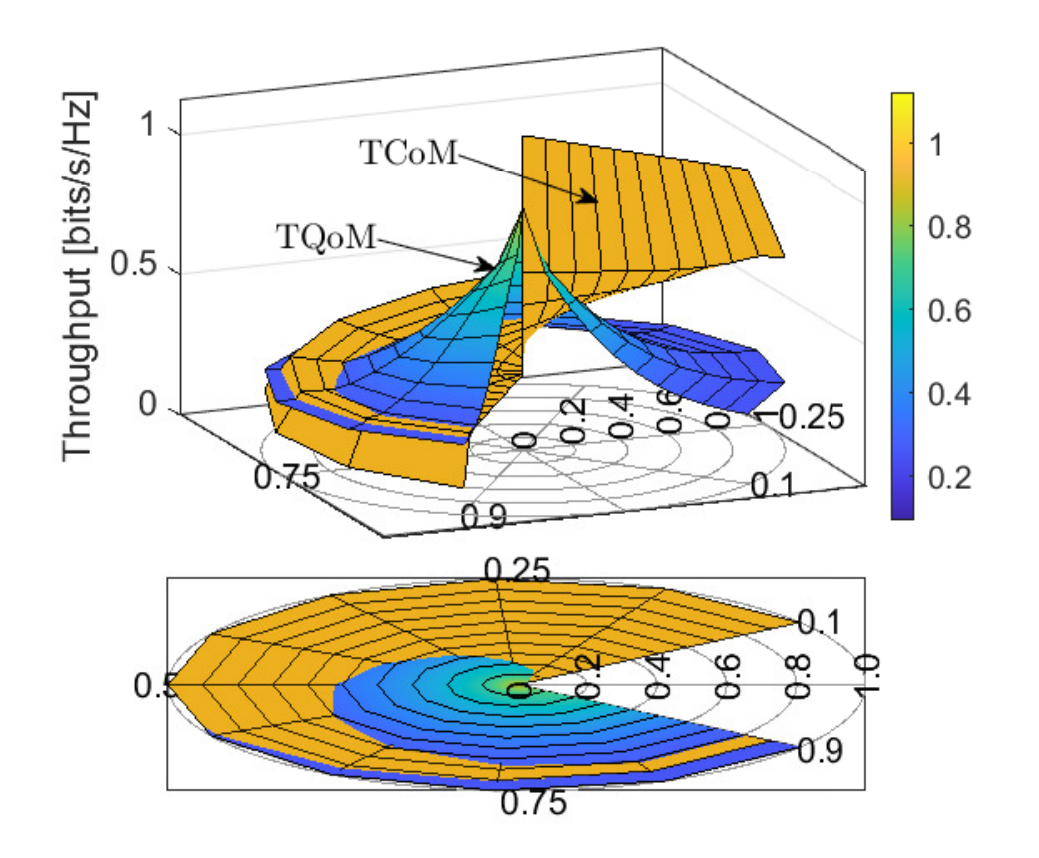}
	\caption{ Network sum-throughput versus time-switching ratio $\alpha$ and probability of EH (i.e., $\rho$). The lower figure represents the projection of the upper figure to the 2D Cartesian plane.}
	\label{fig_throughput_3D_vs_EH_Ratio}
\end{figure}

\begin{figure}[!t]
    \centering
	\adjincludegraphics[width=\linewidth,valign=T]{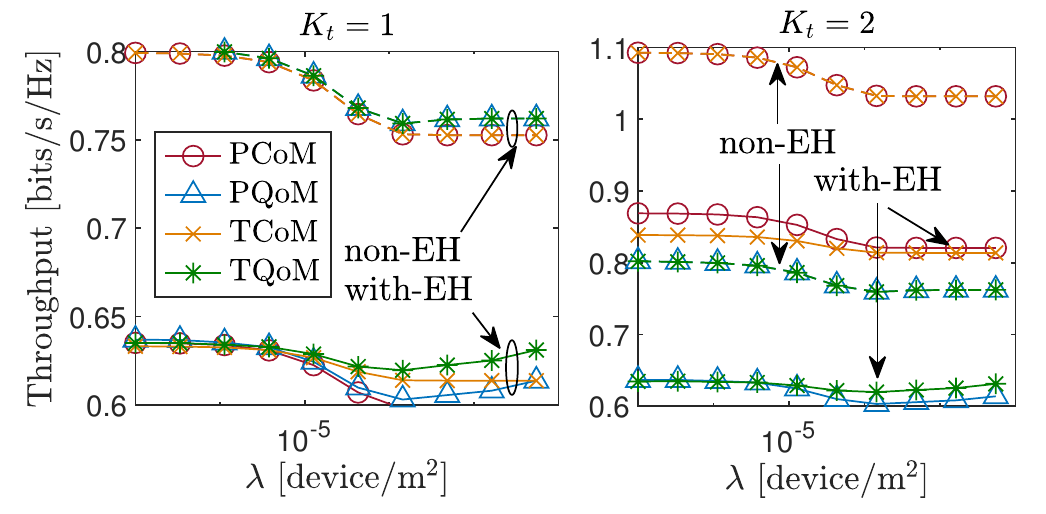}
	\caption{Sum-throughput [bits/s/Hz] versus device density where $P_0 = 0$ [dBm].}
	\label{fig_throughput_vs_density}
\end{figure}

\begin{figure}[!t]
    \centering
	\adjincludegraphics[width=\linewidth,valign=T]{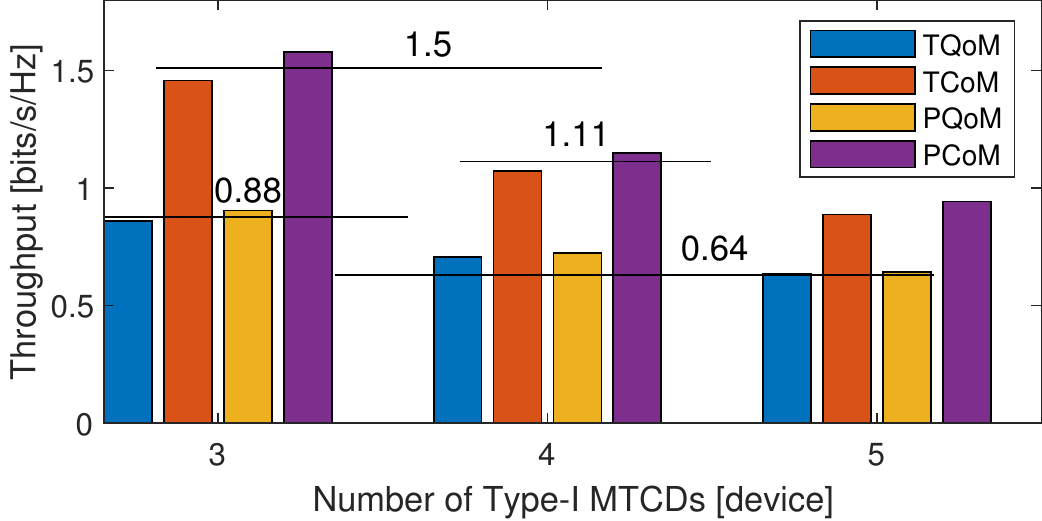} \caption{Maximum sum-throughput [bits/s/Hz] where up to $3$ $\device^\two$s can be served simultaneously.}
    \label{fig_throughput_vs_M}
\end{figure}

As shown in Fig. \ref{fig_network_topo}, where 
    $\textstyle {\bf T}_1 \!=\! \left(\begin{smallmatrix}
        200 & 200 & 200 \\
        100 & 100 & 100 \\
          3 &   2 &   1
    \end{smallmatrix}\right)$, 
    $\textstyle {\bf T}_2 \!=\! \left(\begin{smallmatrix}
        200 & 100 & 100 \\
        100 & 50 & 50 \\
         3 &  2 &  1
    \end{smallmatrix}\right)$,
    $\textstyle {\bf T}_3 \!=\! \left(\begin{smallmatrix}
        50 & 50 \\
        25 & 25 \\
        2 &  2
    \end{smallmatrix}\right)$,
    ${\textstyle {\bf T}_4 \!=\! \left(\begin{smallmatrix}
        200 & 200 & 200 \\
        100 & 100 & 100 \\
         2 &  2 &  2
    \end{smallmatrix}\right)}$, and
    $\textstyle {\bf T}_5 \!=\! \left(\begin{smallmatrix}
        200 & 200 & 200 & 200 \\
        100 & 100 & 100 & 100 \\
         2 &  2 &  2 &  2
    \end{smallmatrix}\right)$, we consider different network topologies in this section.
In addition, numerical examples are provided to validate the analysis and provide additional insight into the performance of the proposed multi-user PD-NOMA wireless-powered mMTC network.
Hereafter, unless otherwise stated, we set the simulation parameters as shown in Table~\ref{table_parameters}. 
In addition,
we set the EH ratios as $\alpha_m = 0.2$ and ${\beta_m = 0.8}$. {Moreover, we consider that the EH probabilities are $\rho_2=\cdots =\rho_{M-1} = \rho = 10^{-1}$. The power allocation $p_M$, $p_{t,k}$, $\forall t \in [1,M-1]$, $\forall k \in [1,K_t]$, is adopted from \cite{WanTC2018}.}
    In addition, we consider 50\% of the MRTR with a maximum of 0.75 [bits/s/Hz] of the target transmission rates during the transmission from $\device^\one_1$ to $\device^\one_M$.
    It is noted that Figs. \ref{fig_e2e_OP_vs_P0}-\ref{fig_COM_1p4_2} adopt the network topology ${\bf T}_{1}$ in Fig. \ref{fig_network_topo}, where the number of $\device^\two$s joining the NOMA transmission is $K_t = M-t$ devices where $t \in [1,M-1]$. 
{
    Moreover, Fig. \ref{fig_throughput_2D_vs_EH_Ratio}-\ref{fig_throughput_vs_density} use the network topology ${\bf T}_{2}$, while 
    Fig. \ref{fig_throughput_vs_M} uses ${\bf T}_{3}$, ${\bf T}_4$, and ${\bf T}_{5}$ for the results of $M = 3$, $M = 4$, and $M = 5$, respectively.
}
From Fig. \ref{fig_e2e_OP_vs_P0} to Fig. \ref{fig_throughput_2D_vs_EH_Ratio}, we compare the results of QoM and CoM schemes with and without being powered by the probabilistic EH architectures to the results of conventional non-regenerative relaying (CNRR) scheme, which consists only non-regenerative relaying transmission from $\device^\one_1$ to $\device^\one_M$ while neglecting the $\device^\two$s. 

Fig. \ref{fig_e2e_OP_vs_P0} provides a comparison between the e2e OP of regenerative and non-regenerative relaying under different values of the fixed transmission power $P_0$ [dBm].
It is observed that the analytical results and the simulation results match, which validates our analysis.
    By increasing $P_0$ [dBm], the e2e OP at $\device^\one_M$ eventually decreases, which improves network reliability.
In the case of no EH, the proposed transmission schemes, which are represented by the yellow curves (QoM/CoM w/o EH), exhibit higher OP than the CNRR scheme. 
    The reason behind this is that the proposed transmission schemes utilize a portion of $P_0$ to serve the $\device^\two$s, whereas the CNRR scheme uses all the power to relay information to the $\device^\one$s and neglects the $\device^\two$s.
In the context of EH-enabled transmissions, the TCoM/TQoM schemes outperform the PCoM/PQoM schemes in terms of reliability. 
    {When jointly adopting the proposed QoM and CoM schemes, the traditional TS-based and PS-based EH architectures in \cite{Nasir2013TWC} requires more transmission power to reach the same outage performance in comparison to that of  the proposed schemes.}

Fig. \ref{fig_throughput_vs_P0} presents the sum-throughput of the CoM and QoM schemes with and without EH operations, {in which the dotted upper bounds in the throughput is $\frac{1}{M-1}(R_M + \sum_{t=1}^{M-1} \sum_{\kappa \in {\cal K}_t} {R_{t,\kappa}})$.}
    The solid black curve shows an upper limit of $0.11$ [bits/s/Hz] in the sum-throughput as $P_0$ reaches $0$ [dBm].
{Compared to the CNRR scheme, the proposed schemes can increase the sum-throughput by an order of magnitude of one, i.e., $OM_1 = OM_2 = 1$, at $P_0 = 0$ [dBm].
    This is because the CNRR system utilizes all of the power to relay information to $\device^\one_M$ while ignoring the $\device^\two$s, and thus it does not gain additional throughput by serving extra devices, unlike the proposed schemes.
As $P_0$ increases beyond $0$ [dBm], more power is wasted for relaying operations, i.e., increasing $X_{t+1} - \tau^{[00]}_{M}$, rather than serving additional $\device^\two$s. 
    Meanwhile, the proposed schemes can utilize that wasted power to serve the $\device^\two$s, thus increasing both connectivity and sum-throughput.}

In Fig. 5, we compare the EE of our proposed schemes with the traditional non-regenerative relaying multihop-based mMTC. 
    The EE is given by ${\sf EE} = \frac{ BW \times {\cal T}_{\Sigma} }{ P_{\rm tol} }$ [Mbits/Joule] \cite{HashemiTGCN2022}, where $P_{\rm tol}$ [W] is the average total power consumed in the transmission from $\device^\one_{1}$ to $\device^\one_{M}$.
In the proposed probabilistic EH architectures, we obtain
\begin{align}
    P_{\rm tol} = \mathbb{E}[ P_0 \sum_{j=1}^{M-1}{\mathbb{I}_j} ] = P_0 \sum_{j=1}^{M-1} {\rho^{[0]}_{t} }.
\end{align}
The EED in Fig. \ref{fig_COM_1p4_2} is determined as
$ {\sf EED} = {\sf EE} - {\sf EE}_{0}$ [Mbits/Joule], where ${\sf EE}_{0}$ is the EE of the non-regenerative relaying with CoM/QoM scheme. The higher the EED is, the more energy efficient the BTEH/BPEH architecture is compared to non-regenerative relaying. 
    Among the proposed schemes, the TQoM has the highest EED, which can reach to $2.5$ [Gbit/Joule].
    This behavior is because the e2e OP at the MTCDs is relatively low when $P_0$ is relatively high, e.g., when $P_0 = 0$ [dBm] in Fig. \ref{fig_throughput_vs_P0}, and the sum-throughput approaches an upper-bound value of $\frac{1}{M-1}(R_M \sum_{t=1}^{M-1} \sum_{\kappa \in {\cal K}_t} {R_{t,\kappa}})$ where continuing to increase $P_0$ lowers the EE, which in turn reduces the EED. 
Since the EE is a decreasing function of $P_{\rm tol}$, as we further increase $P_0$ while $R_{\Sigma}$ remains constant, the EED decreases to zero, which depicts that ${\sf EE} = {\sf EE}_0$.  


Fig. \ref{fig_throughput_2D_vs_EH_Ratio} illustrates the sum-throughput [bits/s/Hz] versus the EH ratios, where $\alpha_i = \alpha$ and $\beta_i = \beta$, $\forall i \in [2,M-1]$.
    In this figure, as the plotted curves move further away from the center point, the sum-throughput of the illustrated schemes increases and vice versa.
Compared to the CNRR scheme presented by the solid black curve, we observe a gain in the sum-throughput of nearly $0.61$ [bits/s/Hz] for the PQoM scheme and $0.75$ [bits/s/Hz] for the non-generative CoM scheme, i.e., the cyan curve.
    Note that varying the EH ratios, i.e., $\alpha$ and $\beta$, does not affect the sum-throughput of non-regenerative relaying using QoM and CoM schemes.
    The reason behind such performance gaps is that the proposed network benefits from serving $\device^\two$s while maintaining relaying information to $\device^\one_M$, whereas the CNRR scheme represents the sum-throughput when $\device^\two$s are ignored.

To further examine the BTEH-powered CoM and QoM schemes, Fig. \ref{fig_throughput_3D_vs_EH_Ratio} illustrates the sum-throughput versus EH ratio $\alpha$ and the probability of EH ($\rho$).
    When $\rho$ increases from $\rho = 0$ to $\rho = 0.4$, the sum-throughput of the TQoM scheme drops from $0.86$ [bits/s/Hz] to about $0.45$ [bits/s/Hz], whereas that of the TCoM scheme is hardly affected and remains around $1.12$ [bits/s/Hz].
However, as $\rho$ increases beyond $0.8$, the TQoM scheme eventually outperforms the TCoM scheme due to less block time consumed for EH, such as ${\alpha \!<\! 0.4}$.
    When more time is consumed for EH, such as ${\alpha \!\ge\! 0.75}$, the TQoM scheme tends to achieve a higher throughput than the TCoM scheme.

In Fig. \ref{fig_throughput_vs_density}, we observe the network sum-throughput for different device density [$\rm{device/m}^2$] values. 
    The left-hand side figure shows that the QoM schemes outperform CoM schemes when the network only serves one $\device^\two$ during each time slot. %
The fundamental reason for this is that QoM schemes schedule the nearest $\device^\two$, but the CoM schemes choose one $\device^\two$ without considering its channel condition. 
    As a result, the received power at the selected $\device^\two$ using QoM schemes is substantially higher than using CoM schemes, and hence the sum-throughput of the QoM schemes is larger than that of the CoM schemes.
However, as shown in the right-hand side figure, when there are more devices served per transmission, e.g., $K_t = 2$, the sum-throughput using the CoM schemes further increases and even outperforms that of the QoM schemes.
    Under non-regenerative relaying, as the network becomes much more populated with MTC devices, the network tends to promote NOMA transmission more frequently, which can further improve the sum-throughput.
We observe that as the network is more populated by $\device^\two$s, i.e., larger values of $\lambda$, the sum-throughput of the PCoM scheme decreases until reaching a floor level.
    The reason for this performance loss while adopting the BPEH architecture is the reduction in the portion of the power utilized for information processing to serve the $\device^\two$s without lowering the target transmission rates. 
    

In Fig. \ref{fig_throughput_vs_M}, we illustrate the maximum sum-throughput versus the number of $\device^\one$s under i.i.d. setting.
    Specifically, we consider that, when using the TCoM or PCoM
schemes, each $\device^\one$ can serve from $K_t = 1$ to $K_t = 3$ $\device^\two$s during each time slot.
With $M = 3$, both $\device^\one_1$ and $\device^\one_2$ serving $K_1 = 1$ and $K_2 = 1$ $\device^\two$, respectively, is considered as one network configuration. 
    When $K_t \le 3$, there would be a total of ${3}^{M-1}$ network configurations.
Then, we consider the maximum sum-throughput obtained from all ${3}^{M-1}$ configurations, which is shown in Fig. \ref{fig_throughput_vs_M}.
    In general, $\device^\one_1$ eventually requires more time slots to relay information to $\device^\one_M$ as $M$ increases, thus decreasing the network sum-throughput.
By increasing $M$ from $3$ to $4$, the maximum sum-throughput using TCoM and PCoM schemes decreases from $1.50$ [bits/s/Hz] to $1.11$ [bits/s/Hz], which corresponds to a $26\%$ reduction in the sum-throughput.
    The maximum sum-throughput using the TQoM and PQoM schemes also decrease as $M$ increases to $4$, from nearly $0.88$ [bits/s/Hz] to $0.64$ [bits/s/Hz], which corresponds to a $27\%$ reduction in the sum-throughput. 
One reason for this behavior is that as $M$ increases, $\device^\one_1$ takes additional time slots to relay information to $\device^\one_M$, thus reducing the sum-throughput of every network configuration and eventually decreasing the maximum sum-throughput.

\section{Conclusions}

In this paper, we proposed two transmission schemes, namely CoM and QoM schemes, for multi-user PD-NOMA probabilistic wireless-powered mMTC networks. 
    Specifically, the proposed schemes can serve two types of MTCDs with different communication priorities. 
Based on a Bernoulli random process-based EH, the proposed schemes are self-sustaining. 
    We considered that the active and inactive MTCDs appear following a homogeneous Poisson point process over the network area. 
We evaluated the performance of the proposed schemes in terms of sum-throughput and derived closed-form expressions for the e2e OP, sum-throughput, and EE. 
    Our results showed that significant gains in the network sum-throughput and EE can be obtained by operating the proposed schemes. 
The results also revealed that employing NOMA in low device density networks hinders a slight sum-throughput loss using the BTEH protocol, whereas a slight improvement in the sum-throughput is observed for the BPEH protocol. 
    We also pointed out the drawbacks of the proposed schemes, which is their maximum achievable throughput deteriorates as the number of relaying devices increases.

\appendices
\section{Proof of Lemma \ref{lem:1}} \label{apd_a}

\subsection{Derivation of the CDF of $\xi_{\tau,t}$ in \eqref{eq_F_xi}} \label{apx:A1}
The CDF, $F_{\xi_{\tau,t}}(\xi)$, can be expressed as
{\allowdisplaybreaks
\begin{align}
&F_{\xi_{\tau,t}}(\xi)
	= \Pr\bigg(
		\Omega_\tau \phi_\tau
		\prod\limits_{i=\tau+1}^{t} {\Omega_i \phi_i} < \xi
	\bigg) \nonumber \\
&
	= 1 - \underbrace{\int_0^\infty \cdots \int_0^\infty}\limits_{t-\tau~\text{folds}}
		\bar{F}_{\phi_\tau}\bigg(
			\bigg(
				\prod_{i=\tau}^t \Omega_i 
				\prod_{i=\tau+1}^{t-1} y_i
			\bigg)^{-1} \frac{\xi}{y_t}
		\bigg) 
\nonumber\\
    &\qquad\qquad\qquad\qquad\qquad\times
    \prod\limits_{i=\tau+1}^{t} 
			f_{\phi_{i}}(y_{i}) {\rm d} y_{i}. \label{apdA_Fxi_Integral}
\end{align}}

Using the identity $e^{-x} = G^{1,0}_{0,1}\left[ x  |  0 \right] = G^{0,1}_{1,0}\left[ x^{-1}  |  1 \right]$, we rewrite the PDF and CCDF of $\phi_\tau$ in terms of Meijer-G functions. The first fold of \eqref{apdA_Fxi_Integral} in terms of $y_t$, denoted by $I_1(\xi,y_{t-1})$, can be expressed as
\begin{align}
&I_1(\xi,y_{t-1}) 
    = \!\!\! 
    \int\limits_0^\infty
	\frac{ G^{1,0}_{0,1}\left[
		\left.
			 \frac{\phi}{\ell(l_t)}
		\right\vert 0
	\right] }{\ell(l_t)}
\nonumber\\
&\qquad\qquad\quad\times
	G^{0,1}_{1,0}\bigg[
		\bigg( 
			\prod_{i=\tau}^t \Omega_i \!\!\!
			\prod_{i=\tau+1}^{t-1} y_i
		\bigg) \frac{\ell(l_\tau)}{\xi} y_t
		\bigg\vert 1
	\bigg] {\rm d} y_t
\nonumber\\
&\qquad\quad
    = G^{0,2}_{2,0}\bigg[
		\frac{\ell(l_\tau) \ell(l_t)}{\xi}
		\prod_{i=\tau}^{t} {\Omega_i} \!\!\!
		\prod_{i=\tau+1}^{t-1} y_i
		\bigg\vert 0,1
	\bigg].
\label{eq:A6}
\end{align}

Repeating similar steps for the subsequent integrals, we then obtain $F_{\xi_{\tau,t}}(\xi)$ in \eqref{eq_F_xi}. 

\subsection{Derivation of the PDF of ${\xi_{\tau,t}}$ in \eqref{eq_f_xi}} \label{apx:A2}

The proof of \eqref{eq_gamma_nq_m1_nk_end} is quite similar to that of \eqref{eq_gamma_m1_nk_end}. It should be noted that by taking the derivatives of \eqref{eq:A6}, the PDF of $\xi_{\tau,t}$ can be rewritten as
\begin{align}
f_{\xi_{\tau,t}}(\xi) 
    &= \bigg( 
		\prod_{i=\tau}^t \frac{1}{\Omega_i}
	\bigg) 
    \underbrace{\int_0^\infty \cdots \int_0^\infty}\limits_{t-\tau~\text{folds}}
    \prod\limits_{i=\tau+1}^{t} 
		\frac{ f_{\phi_{i}}(y_{i}) }{ y_i }
\nonumber\\
&\times
	f_{\phi_\tau}\bigg(
		\bigg( 
			\prod_{i=\tau}^t \Omega_i 
			\prod_{i=\tau+1}^{t-1} y_i
		\bigg)^{-1} \frac{\xi}{y_t}
	\bigg) {\rm d} y_{i}.
\label{eq:A9}
\end{align}

Rewriting $f_{\phi_\tau}(x)$ and $\frac{f_{\phi_i}(y_i)}{y_i}$ in terms of the Meijer G-function, we obtain
\begin{align}
\frac{f_{\phi_i}(y_i)}{y_i} 
	= 	\frac{1}{\ell(l_i)}
	G^{1,0}_{0,1}\bigg[
		\frac{\phi}{\ell(l_i)}
		\bigg\vert -1
	\bigg],~y_i > 0,
\label{eq:A10}
\end{align}
then the first fold of the above integral, denoted by $J_1(\xi)$, can be obtained as
\begin{align}
J_1(\xi) =
	\frac{1}{\ell(l_\tau) \ell(l_t)}
    G^{0,2}_{2,0}\bigg[
		\frac{\ell(l_\tau) \ell(l_t) }{\xi}
		\prod_{i=\tau}^{t} \Omega_i \!\!\!
		\prod_{i=\tau+1}^{t-1} y_i 
		\bigg\vert 1,1
	\bigg].
\label{eq:A11}
\end{align}

Substituting \eqref{eq:A11} into \eqref{eq:A9} and repeating the steps in \eqref{eq:A10} and \eqref{eq:A11} for the remaining integrals, we then obtain $f_{\xi_{\tau,t}}(\xi)$ as shown in \eqref{eq:A9}, which concludes the~proof.

\section{Proofs of \eqref{eq_F_Xm_0} and \eqref{eq_F_Xm_1}} \label{apendix_proof_lemma_X_fin}

Using the representation of the transmission power $P_t$ provided by \eqref{eq_unified_P_t} and then applying the total probability theorem, the CDF of $X_{t+1}$, $\forall t \in [1,M-1]$, can be expressed as
\begin{align} 
&F_{X_{t+1}}(x) 
	= 1 - \Pr\Bigg[
		 \frac{P_0 \phi_{t+1}}{\sigma^2} \ge x, 
		 \big[ \mathbbm{I}_{t} = 0 \big]
	\Bigg]
\nonumber\\
&\quad
    -  \sum_{\tau = 1}^{t-1}
	\Pr\Bigg[
		 \frac{\phi_{t+1} \prod_{i=\tau+1}^{t} \Omega_i \phi_i}{\sigma^2} \ge x, \big[ \mathbbm{I}_{\tau} = 0 \big], 
\nonumber\\
&\qquad\qquad\qquad\qquad\qquad\quad
		 \bigcap_{j=\tau+1}^{t} \big[ \mathbbm{I}_{j} = 1 \big]
	\Bigg].  \label{eq_F_X} 
\end{align}
The first probability can be derived as
\begin{align}
\bar{F}_{X_{t+1}^{[0]}}(x)
    =   \rho^{[0]}_t 
    \!\!\!\!
    \int\limits_{\phi > {x}\big/{\bar{\gamma}_0} }
    \!\!\!\!
	f_{\phi_{t+1}}(\phi)d\phi \mathop{=}\limits^{(a)} \eqref{eq_F_Xm_0}
\label{eq_F_Xm_proof_final}
\end{align}
where $(a)$ adopts $f_{\phi_{t+1}}(\phi) = \frac{e^{-x/\ell(l_{t+1})}}{\ell(l_{t+1})}$ and some straightforward mathematical manipulations.

Since $\mathbbm{I}_0$, $\mathbbm{I}_1$, $\dots$, and $\mathbbm{I}_M$ are mutually independent Bernoulli trials, and since $\prod_{i=\tau+1}^{t} \Omega_i \phi_i = \xi_{\tau+1,t}$, whose PDF is provided in Lemma \ref{lemma_TypeI_BTEH_xi_tau_t}, the complementary CDF $\bar{F}_{X_{t+1}^{[1]}}(x)$
can be derived in an integral-form as
\begin{align}
\bar{F}_{X_{t+1}^{[1]}}(x)
	&= \sum_{\tau = 1}^{t-1}
	\Bigg[
		{\rho}^{[0]}_\tau
		\prod_{j=\tau+1}^{t} { \rho^{[1]}_j }
	\Bigg] 
\nonumber\\
    &\quad\times
    \int\limits_0^\infty
	\bar{F}_{\phi_{t+1}}\left(
		\frac{x}{\bar{\gamma}_0 \xi}
	\right) f_{\xi_{\tau+1,t}}(\xi) {\rm d} \xi.
\label{eq:A14}
\end{align}

By substituting the CDF of $\phi_{t+1}$ and the PDF of $\xi_{\tau+1,t}$ provided in Lemma \ref{lemma_TypeI_BTEH_xi_tau_t}, the probability $\bar{F}_{X_{t+1}^{[1]}}(x)$ can be further expressed as
\begin{align}
&\bar{F}_{X_{t+1}^{[1]}}(x)
	= \sum_{\tau=1}^{t-1}
	\frac{ {\rho}^{[0]}_\tau
		\prod_{j=\tau+1}^{t} { \rho^{[1]}_j } }{\bar{\xi}_{\tau+1,t}} 
\nonumber\\
&\quad\times
\int\limits_0^\infty
		G^{0,1}_{1,0}
		\Bigg[
			\frac{\bar{\gamma}_0 \ell(l_{t+1})}{x}  \xi
			\Bigg\vert 1
		\Bigg] 
	G^{t-\tau,0}_{0,t-\tau}
	\left[
	\left.
		\frac{\xi}{\bar{\xi}_{\tau+1,t}}
	\right\vert
		{\bf 0}_{t-\tau}
	\right] {\rm d} \xi.
\label{eq:A15}
\end{align}

Since the integration over a product of two Meijer G-functions is also a Meijer G-function, we then obtain $\bar{F}_{X_{t+1}^{[1]}}(x)$ in \eqref{eq_F_Xm_1}.
    Substituting \eqref{eq_F_Xm_proof_final} and the result of \eqref{eq:A15} into \eqref{eq_lemma_FX_sim}, we obtain the CDF of $X_{t+1}$.
This completes the proof of \eqref{eq_F_Xm_0} and \eqref{eq_F_Xm_1}.

\section{Proof of \eqref{eq:CCDF_Y_tk}} \label{apendix_proof_lemma_TypeII_CoM_Ytk}

Using $P_t$ in \eqref{eq_unified_P_t}, the CDF of $Y_{t,k}$ can be expressed as
\begin{align}
F_{Y_{t,k}}(y) 
    &= 1 - \Pr\left\{
	\bar{\gamma}_0 \varphi_{t,k} \ge y, 
	\big[
	\mathbbm{I}_t = 0
	\big]
	\right\}
\nonumber\\
&\quad
	- \sum_{\tau=1}^{t-1}
	\Pr\bigg\{
	\varphi_{t,k} 
	\ge \frac{y}{\bar{\gamma}_0 \prod_{{i = \tau+1}}^{t}{ \Omega_i \phi_i }}, 
\nonumber\\
&\qquad\qquad
	\big[ \mathbbm{I}_\tau = 0 \big],
        \bigcap_{{j=\tau+1}}^{t} { \big[ \mathbbm{I}_j = 1 \big] }
	\bigg\}.
\end{align}

The derivation of the first probability in the above equation is straightforward. For the second probability, denoted by $\bar{F}^{[1]}_{t,k}$, it is noted that $\xi_{\tau+1,t} = \prod_{{i = \tau+1}}^{t}{ \Omega_i \phi_i }$. In addition, since $\mathbbm{I}_\tau$, $\mathbbm{I}_{\tau+1}$, $\dots$, $\mathbbm{I}_t$ are mutually independent, the above expression can be rewritten as
\begin{align}
\bar{F}_{Y_{t,k}^{[1]}}(y)
&= 	\sum_{\tau = 1}^{t-1} { \bigg[ 
	    {\rho}^{[0]}_\tau \prod_{{j = \tau+1}}^{t}{ \rho^{[1]}_j }
	\bigg] }
\nonumber\\
&\quad\times
	\underbrace{
		\int_0^\infty
		\bar{F}_{\varphi_{t,k}}\left(
		\frac{y}{\bar{\gamma}_0 \xi}
		\right) f_{\xi_{\tau+1,t}}(\xi) {\rm d} \xi
	}\limits_{J_{t,k}} 
    .
\label{eq_apxB_2}
\end{align}

In order to solve the above integral, we rewrite the CCDF $\bar{F}_{\varphi_{t,k}}$ using the Meijer G-function as 
\begin{align}
\bar{F}_{\varphi_{t,k}}(\varphi) 
	&= \frac{2}{\epsilon}
	\chi^{[0]}_{k} 
	G^{1,1}_{1,2}
	\left[
	\frac{\big(
		\frac{k}{K_t}
		\big)^\epsilon}{\ell(r_t)}
	\varphi
	\left\vert 
	\begin{array}{cc}
		1 - \frac{2}{\epsilon} \\
		0, -\frac{2}{\epsilon} 
	\end{array}
	\right.
	\right]
\nonumber\\
    &
    - \frac{2}{\epsilon}
	\chi^{[1]}_{k} 
	G^{1,1}_{1,2}
	\left[
	\frac{\big(
		\frac{k-1}{K_t}
		\big)^\epsilon}{\ell(r_t)}
	\varphi
	\left\vert 
	\begin{array}{cc}
		1 - \frac{2}{\epsilon} \\
		0, -\frac{2}{\epsilon} 
	\end{array}
	\right.
	\right],
	\label{eq_CDF_varphi_mk}
\end{align}
where $\varphi > 0$ and $\chi^{[i]}_{k} \triangleq \frac{ (k-i)^2 }{ k^2 - (k-1)^2 }$.

Substituting the above equation and $f_{\xi_{\tau+1,t}}(\xi)$ in \eqref{eq_f_xi} into \eqref{eq_apxB_2} and applying the identity \cite[Eq. (7.811.1)]{Gradshteyn2007}, we obtain the result in \eqref{eq:CCDF_Y_tk}. This complete the proof of \eqref{eq:CCDF_Y_tk}.

\section{Proof of \eqref{eq_CDF_varphi_t}}
\label{apx:B}

The PDF of the nearest distance, denoted by $f_{d_t}(r)$, can be obtained as
\begin{align}
f_{d_t}(r)
	=   \frac{2 \pi \lambda_t}{1-e^{-\pi \lambda_t r_t^2}} r 
		e^{ -\pi \lambda_t r^2 },~0 \le r \le r_t.
\label{eq:A16}
\end{align}

Subsequently, the CDF $F_{\varphi_t}(\varphi)$ can be expressed in an integral-form as
\begin{align}
&F_{\varphi_t}(\varphi)
	= 1 - \int_0^{r_t}
		\exp\left( -\frac{r^\epsilon}{\mathcal{L} d_0^\epsilon} \varphi \right)
			f_{d_t}(r) {\rm d} r. 
\nonumber \\
	&\mathop{=}\limits^{(a)}
	1 - \frac{\pi \lambda_t r_t^2}{1-e^{-\pi \lambda_t r_t^2}}
	\frac{2}{\epsilon}
	\int_0^{1} u^\frac{2-\epsilon}{\epsilon}
\nonumber\\
    &\qquad\qquad\times
	\exp\left( -u \frac{\varphi}{\ell(r_t)} - \pi \lambda_t r_t^2 u^\frac{2}{\epsilon} \right) {\rm d} u,
\label{eq:A18}
\end{align}
in which $(a)$ is obtained by substituting \eqref{eq:A16} into the first integral and then applying the change of variable $u \leftarrow r^\epsilon$. 

To the best of the authors knowledge, the above equation is intractable to derive in closed-form. 
    In order to provide a tractable expression for $F_{\varphi_t}(\varphi)$, we adopt the curve fitting method that matches the CDF of $\varphi_t$ to that of a Singh-Maddala distributed RV with scale $\mu_t$, and shapes $\theta_t$ and $m_t$. 
In other words, \eqref{eq:A18} can be fitted to
\begin{align}
F_{\varphi_t}(\varphi) 
	= 1 - \left( 
	    1+\frac{\varphi^{\theta_t}}{\mu_t^{\theta_t}} 
    \right)^{-m_t},~\varphi > 0.
\label{eq_vphi_t}
\end{align}

Finally, we use the identities \cite[Eq. (2.9.6)]{Kilbas2004} and \cite[Eq. (2.1.3)]{Kilbas2004} to rewrite \eqref{eq_vphi_t} in terms of the Fox H-function. This completes the proof of \eqref{eq_CDF_varphi_t}.

\section{Proof of \eqref{eq_F_Z_t}} \label{apendix_proof_lemma_Z_fin}
Recalling that $Z_t = P_t \varphi_t/\sigma^2$ is a function of $\varphi_t$. Consequently, it is mandatory to examine the CDF of $\varphi_t$ before that of $Z_t$. Thus, we introduce the following Lemma.

Invoking $P_t$ in \eqref{eq_unified_P_t}, the CDF of $Z_t$ where $1 \le t \le M-1$ can be expressed as
\begin{align}
&F_{Z_t}(z) 
    = 1 - \Pr\left[
        \varphi_t \ge \frac{z}{ \bar{\gamma}_0 },
        \big[ 
            \mathbbm{I}_{t} = 0
        \big]
    \right]
\nonumber\\
&
    - \sum_{\tau=1}^{t-1}
    \Pr\left[
        \varphi_t \ge \frac{z}{ \bar{\gamma}_0 \xi_{\tau+1,t} },
        \big[ 
            \mathbbm{I}_{\tau} = 0
        \big],
        \bigcap_{{j=\tau+1}}^{t} { [ \mathbbm{I}_j = 1 ] }
    \right].
\end{align}

Since $\varphi_t$ and $\mathbbm{I}_t$ are independent, we obtain
\begin{align}
&F_{Z_t}(z) 
    = 1 - \rho^{[0]}_t
    \Pr\bigg[
        \varphi_t \ge \frac{z}{ \bar{\gamma}_0 }
    \bigg] 
\nonumber\\
&
    - \sum_{\tau=1}^{t-1}
    \bigg[ 
        \rho^{[0]}_\tau
        \prod_{j=\tau+1}^{t} {\rho^{[1]}_j}
    \bigg]
    \Pr\bigg[
        \varphi_t \ge \frac{z}{ \bar{\gamma}_0 \xi_{\tau+1,t} }
    \bigg].
\label{eq_Zt_proof_2}
\end{align}

The second probability, denoted by $J'_t$, can be expressed as
\begin{align}
J'_t = \int\limits_0^\infty 
    \bar{F}_{\varphi_t}\left(
        \frac{z}{\bar{\gamma}_0 \xi}
    \right) 
    f_{\xi_{\tau+1,t}}(\xi) {\rm d} \xi.
\label{eq_Jt_proof_1}
\end{align}

It should be noted that the PDF $f_{\xi_{\tau+1,t}}(\xi)$ can be rewritten using the Fox H-function as
\begin{align}
&f_{\xi_{\tau+1,t}}(\xi)
    = \frac{1}{\bar{\xi}_{\tau+1,t}}
\nonumber\\&\quad\times
    H^{t-\tau+1,0}_{0,t-\tau+1}
    \left[ 
        \left.
        \frac{\xi}{\bar{\xi}_{\tau+1,t}}
        \right\vert 
        \underbrace{(0,1),\dots,(0,1)}\limits_{t-\tau~{\rm terms}}  
    \right]
    ,~\xi > 0.
\label{eq_xi_foxH}
\end{align}

Substituting \eqref{eq_CDF_varphi_t} and the above equation into \eqref{eq_Jt_proof_1}, $J'_t$ can then be expressed as
\begin{align}
&J'_t = \int\limits_0^\infty
    \frac{1}{\Gamma(m_t)} 
    H^{1,1}_{1,1}
    \Bigg[
        \bigg( 
    	    \frac{\bar{\mu_t \gamma}_0}{z} \xi
    	\bigg)^{\theta_t}
    	\bigg\vert
    	\begin{array}{cc}
    		(1,1) \vspace{-5pt} \\ (m_t,1) 
    	\end{array}
    \Bigg] 
\nonumber\\&\qquad\times
    \frac{1}{\bar{\xi}_{\tau+1,t}} 
    H^{t-\tau+1,0}_{0,t-\tau+1}
    \Bigg[ 
        \frac{\xi}{\bar{\xi}_{\tau+1,t}}
        \bigg\vert
        ({\bf 0}, {\bf 1})_{t-\tau}
    \Bigg]
    {\rm d} \xi.
\end{align}

Finally, we use the identity \cite[Eq. (2.8.4)]{Kilbas2004} to solve the above integral and substitute the resulted expression into \eqref{eq_Zt_proof_2}. This completes the proof of~\eqref{eq_F_Z_t}.

\section{Proof of Lemma \ref{lem_asymp_X}} \label{apx_proof_F}
When $\bar{\gamma}_0 \to \infty$, we can approximate $\bar{F}_{X_{t+1}^{[0]}}(x)$ as
\begin{align}
\bar{F}_{X_{t+1}^{[0]}}(x)
    \to \rho^{[0]}_{t} - \rho^{[0]}_{t} \frac{x}{\bar{\gamma}_0 \ell(l_{t+1})}.
\label{eq_X0_asymp}
\end{align}

Let us denote the Meijer G-function in \eqref{eq_F_Xm_1} as $I_{\tau+1,t}(x;\ell)$, using \cite[Eq. (9.301)]{Gradshteyn2007}, we obtain that
$
{I_{\tau+1,t}(x;\ell) =
    \frac{1}{2\pi i}
    \oint_{\gamma} F(s)
    {\rm d} s}
$,
where $F(s) \triangleq
    \Gamma^{t-\tau}(1+s) \Gamma(s) 
    ( \frac{ x }{ \bar{\gamma}_{0} \ell \bar{\xi}_{\tau+1,t} } )^{-s} $ and $\gamma$ is a suitable line contour, in which all the poles of $\Gamma^{t-\tau}(1+s) \Gamma(s)$ lie on the left of $\gamma$. 
    As $\bar{\gamma}_{0} \to \infty$, we apply the Cauchy's Residual theorem to approximate $I_{\tau+1,t}(x;\ell)$ by finding the residuals of $F(s)$ at the simple pole $s = 0$ and the $(t-\tau+1)$-th order pole $s = -1$~as
\begin{align}
& I_{\tau+1,t}(x;\ell)
    \to
    [ s F(s) ]|_{s\to 0}
    +   \frac{1}{(t-\tau)!}
\nonumber\\
    &\qquad~\times
    \bigg[
        \frac{{\rm d}^{t-\tau}}{ {\rm d} s^{t-\tau} }
        (s+1)^{t-\tau+1} F(s) 
    \bigg]\bigg\vert_{s \to -1},~\bar{\gamma}_{0} \to \infty.
\end{align}

By applying the generalized Leibnitz's rule of derivative \cite{SilbersteinRS1991} and after some mathematical manipulations, we obtain
\begin{align}
&I_{\tau+1,t}(x;\ell) 
    \to 1 - \frac{ x }{ \ell \bar{\gamma}_{0} \bar{\xi}_{\tau+1,t} }
    \sum_{n=0}^{t-\tau} \frac{\psi^{(n)}_{t-\tau}}{n!}
    \sum_{r=0}^{t-\tau-n} \frac{(-1)^r}{r!}
\nonumber\\
&\qquad\qquad\qquad\qquad\quad\times
    \bigg( 
        \log \frac{ x }{ \ell \bar{\gamma}_{0} \bar{\xi}_{\tau+1,t} }
    \bigg)^r.
\end{align}

Plugging this approximation and \eqref{eq_X0_asymp} into \eqref{eq_lemma_FX_sim}, we obtain
\begin{align}
&F_{X_{t+1}}(x) 
    \to \rho^{[1]}_{t} 
    -   \sum_{\tau=1}^{t-1} 
        \bigg[ \rho^{[0]}_{\tau} \prod_{j=\tau+1}^{t} \rho^{[1]}_{j} \bigg]
\nonumber\\
&\quad
    + F_t\bigg(x; \frac{1}{\bar{\gamma}_0 \ell(l_{t+1})} \bigg)
    \frac{x}{\bar{\gamma}_0 \ell(l_{t+1})},~
\bar{\gamma}_{0} \to \infty.
\label{eq_X_asymp}
\end{align}

Note that with $\rho^{[1]}_{1} = 0$, we have
$\rho^{[1]}_{t} - \sum_{\tau=1}^{t-1} [ \rho^{[0]}_{\tau} \prod_{j=\tau+1}^{t} \rho^{[1]}_{j}] = 0$, $\forall t$.
This completes the proof of Lemma \ref{lem_asymp_X}.

\balance
\bibliographystyle{IEEEtran}
\bibliography{References}

\end{document}